\documentclass{jpsj3}

\usepackage{txfonts}
\usepackage{bm}
\usepackage{braket}
\usepackage{color}
\usepackage{graphicx}

\newcommand\sgn{\mathop{\mathrm{sgn}}}

\title{Mean-Field Solution of the Weak-Strong Cluster Problem for Quantum Annealing with Stoquastic and Non-Stoquastic Catalysts}

\author{Kabuki Takada$^1$\thanks{takada@qa.iir.titech.ac.jp}, Yu Yamashiro$^{1,2}$, and Hidetoshi Nishimori$^{3,4,5}$}

\inst{
$^1$Department of Physics, Tokyo Institute of Technology, Nagatsuta-cho, Midori-ku, Yokohama 226-8503, Japan\\
$^2$Jij Inc., High tech Hongo Building 1F, 5-25-18 Hongo, Bunkyo-ku, Tokyo 113-0033, Japan\\
$^3$Institute of Innovative Research, Tokyo Institute of Technology, Nagatsuta-cho, Midori-ku, Yokohama 226-8503, Japan\\
$^4$Graduate School of Information Sciences, Tohoku University, Sendai 980-8579, Japan\\
$^5$RIKEN, Interdisciplinary Theoretical and Mathematical Sciences (iTHEMS),
Wako, Saitama 351-0198, Japan
}

\abst{
We study the weak-strong cluster problem for quantum annealing in its mean-field version as proposed by Albash [Phys. Rev. A \textbf{99} (2019) 042334] who showed by numerical diagonalization that non-stoquastic $XX$ interactions (non-stoquastic catalysts) remove the problematic first-order phase transition. We solve the problem exactly in the thermodynamic limit by analytical methods and show that the removal of the first-order transition is successfully achieved either by stoquastic or non-stoquastic $XX$ interactions depending on whether the $XX$ interactions are introduced within the weak cluster, within the strong cluster, or between them. We also investigate the case where the interactions between the two clusters are sparse, i.e. not of the mean-field all-to-all type. The results again depend on where to introduce the $XX$ interactions. We further analyze how inhomogeneous driving of the transverse field affects the performance of the system without $XX$ interactions and find that inhomogeneity in the transverse field removes the first-order transition if appropriately implemented.
}

\begin{document}

\maketitle

\section{Introduction} \label{sec.intro}

It is an interesting and important problem in quantum annealing\cite{Kadowaki1998,Brooke1999,Santoro2002,Santoro2006,Das2008,Morita2008,Hauke2019} in its implementation as adiabatic quantum computing\cite{Farhi2001,Albash2018} whether or not the introduction of non-stoquastic interactions (non-stoquastic catalysts) enhances the performance compared to the case of the traditional formulation without non-stoquasticity. A stoquastic Hamiltonian can be represented as a matrix with non-positive off-diagonal elements in a product basis of local states, and can be simulated classically without the sign problem\cite{Suzuki1976,Bravyi2008}. Introduction of non-stoquasticity into the Hamiltonian makes it difficult to classically simulate the system\cite{Gupta2019}, but it does not necessarily mean a speedup as compared to the case of a stoquastic Hamiltonian.

Numerical studies of finite-size systems indicate that the introduction of a non-stoquastic catalyst increases the success probability in a small subset of problem instances\cite{Crosson2014,Hormozi2017,Albash2019b}. Analytical studies of the $p$-spin model (a mean-field-type $p$-body interacting ferromagnetic system) and the Hopfield model show that a non-stoquastic catalyst is effective to remove the first-order phase transition, which exists in the original stoquastic Hamiltonian, leading to an exponential speedup compared to the traditional stoquastic case\cite{Seki2012,Seoane2012,Seki2015,Nishimori2017}. In a recent paper, Albash\cite{Albash2019} introduced a mean-field version of the weak-strong cluster problem (also known as the large-spin tunneling problem)\cite{Farhi2002,Boixo2016}, which was used to test the possibility of large-scale tunneling effects in the D-Wave quantum annealer\cite{Denchev2016}. Albash showed by numerical diagonalization of small-size systems that a non-stoquastic catalyst introduced between the two clusters in the problem eliminates the first-order transition that exists in the case without non-stoquasticity. He also introduced a geometrically local Hamiltonian for which evidence was provided for a similar phenomenon. Under these circumstances, it is desirable to study more instances analytically toward the goal of understanding when and how non-stoquastic catalysts lead to (or do not lead to) increased performance, in particular given the ongoing efforts to implement non-stoquasticity at the hardware level\cite{Ozfidan2019}.

We have carried out a comprehensive analytical study of the mean-field version of the weak-strong cluster problem formulated by Albash and its generalization to the case with sparse (not all-to-all) interactions between the clusters, the latter being closer to the realistic hardware implementation. We analytically confirm his numerical conclusion that the non-stoquastic catalyst introduced between the clusters with an appropriate amplitude removes the first-order transition. We have further found that the elimination of the first-order transition is possible even with a stoquastic catalyst if it is introduced in an appropriate way. We also study the effects of inhomogeneous driving of the transverse field in the original stoquastic problem, inhomogeneity meaning that the transverse field is driven more quickly in one of the clusters than in the other. We show that this protocol is effective to eliminate the first-order transition. Our results represent a complete solution to the weak-strong cluster problem with stoquastic or non-stoquastic catalysts in the mean-field framework.

This paper is organized as follows. In Section~\ref{sec.wscl}, we solve the weak-strong cluster problem with dense interactions within the clusters and dense or sparse interactions between the clusters. Section~\ref{sec.conc} concludes the paper. Technical details are relegated to Appendices.

\section{Weak-Strong Cluster Problem} \label{sec.wscl}

We define the weak-strong cluster problem with mean-field-type dense interactions within and between the clusters as proposed in Ref.~\citen{Albash2019} and analyze it in Section~\ref{ssec.wscl_mf}. Then the case with sparse interactions between the clusters is solved in Section~\ref{ssec.wscl_sp}.

\subsection{Weak-strong cluster problem with dense interactions between clusters} \label{ssec.wscl_mf}

\begin{figure}
\centering
\includegraphics[scale=0.3]{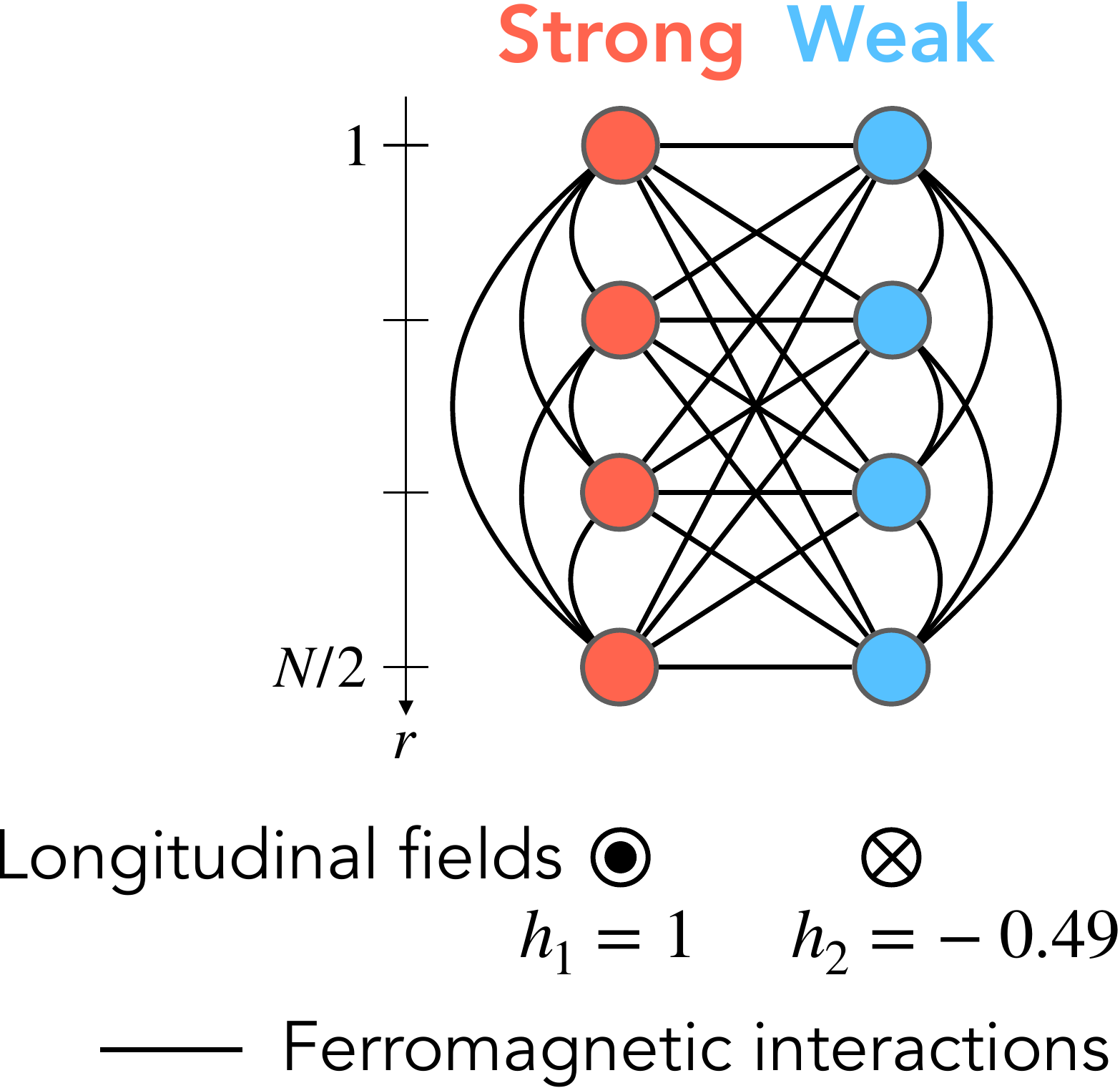}
\caption{(Color online) Weak-strong cluster problem with dense interactions between clusters. The circles in the left (right) side denote the spins in the strong (weak) cluster.} \label{fig.wscl_mf_diagram}
\end{figure}

We first study the weak-strong cluster problem with dense interactions between the two clusters. The model has two subsystems (clusters) with the problem Hamiltonian
\begin{align}
\hat{H}_\mathrm{p} & =-\sum _{r=1}^{N/2} (h_1 \hat{\sigma}_{1r}^z +h_2 \hat{\sigma}_{2r}^z) \notag \\
& \hphantom{{} = {}} -\frac{1}{N} \sum _{r,r'=1}^{N/2} (\hat{\sigma}_{1r}^z \hat{\sigma}_{1r'}^z +\hat{\sigma}_{2r}^z \hat{\sigma}_{2r'}^z +\hat{\sigma}_{1r}^z \hat{\sigma}_{2r'}^z),
\end{align}
where $N$ is the total number of spins (qubits) in the system and $\hat{\bm{\sigma}}_{ar} =(\hat{\sigma}_{ar}^x ,\hat{\sigma}_{ar}^y ,\hat{\sigma}_{ar}^z)$ is the Pauli operator at $(a,r)$, with $a$ ($=1,2$) representing the cluster index and $r=1,\dots ,N/2$ the site index within each cluster. We set the strengths of longitudinal magnetic fields to $h_1 =1$ and $h_2 =-0.49$. Notice that the strong longitudinal field $h_1$ and the weak one $h_2$ point in the opposite directions. We refer to the first subsystem $a=1$ as the strong cluster and the second subsystem $a=2$ as the weak cluster. The structure of the problem is schematically depicted in Fig.~\ref{fig.wscl_mf_diagram}. The ground state of $\hat{H}_\mathrm{p}$ is the eigenstate of $\hat{\sigma}_{1r}^z$ and $\hat{\sigma}_{2r}^z$ with eigenvalues $\sigma _{1r}^z =\sigma _{2r}^z =+1$ (all spins pointing up, to be called `state A'), while a metastable state exists with eigenvalues $\sigma _{1r}^z =+1$ and $\sigma _{2r}^z =-1$ (spins in the strong cluster pointing up and those in the weak cluster pointing down, to be called `state B'). The conventional quantum annealing with a uniform transverse field, in which the Hamiltonian is stoquastic, has a first-order phase transition when state A and state B exchange their (meta)stability, meaning a large-scale spin flip in the weak cluster\cite{Denchev2016}.

Let us construct a quantum annealing Hamiltonian for this problem, generalizing the formulation in Ref.~\citen{Albash2019}. Using the magnetization operators $\hat{\bm{m}}_a =(2/N)\sum _r \hat{\bm{\sigma}}_{ar}$, the Hamiltonian $\hat{H} (s)$ is defined as
\begin{align}
\frac{\hat{H} (s)}{N} & =-\frac{s}{2} \left( h_1 \hat{m}_1^z +h_2 \hat{m}_2^z \right) -\frac{s}{4} \left( (\hat{m}_1^z)^2 +(\hat{m}_2^z)^2 +\hat{m}_1^z \hat{m}_2^z \right) \notag \\
& \phantom{{} = {}} -\frac{1-\gamma _1 (s)}{2} \hat{m}_1^x -\frac{1-\gamma _2 (s)}{2} \hat{m}_2^x \notag \\
& \hphantom{{} = {}} -\frac{s(1-s)}{4} \left(\xi _{11} (\hat{m}_1^x)^2 +\xi _{22} (\hat{m}_2^x)^2 +\xi _{12} \hat{m}_1^x \hat{m}_2^x \right) , \label{eq.wscl_mf_hamiltonian}
\end{align}
where $s\in [0,1]$ denotes the dimensionless time. Suppose that $\gamma _1 (s)$ and $\gamma _2 (s)$ are monotonically increasing functions which satisfy $\gamma _1 (0)=\gamma _2 (0)=0$ and $\gamma _1 (1)=\gamma _2 (1)=1$, and $\xi _{11}$, $\xi _{22}$, and $\xi _{12}$ are constants.

The Hamiltonian $\hat{H} (s)$ consists of the problem Hamiltonian (the first line of Eq.~\eqref{eq.wscl_mf_hamiltonian}) and the driver Hamiltonian (the second and third lines). The strength of the problem Hamiltonian increases in proportion to $s$. The driver Hamiltonian is the sum of time-dependent transverse fields and $XX$ interactions. The strength of the transverse field in each cluster decreases with time. We can achieve inhomogeneous driving of the transverse field by choosing different functions for $\gamma _1 (s)$ and $\gamma _2 (s)$. As for the $XX$ interactions, non-zero $\xi _{ab}$ makes the corresponding term non-vanishing except at the beginning and the end of annealing. When $0<s<1$, the Hamiltonian $\hat{H} (s)$ is non-stoquastic for $\xi _{11} <0$, $\xi _{22} <0$, or $\xi _{12} <0$ and stoquastic for $\xi _{11} ,\xi _{22} ,\xi _{12} \geq 0$.

For the moment, we assume $\gamma _1 (s)=\gamma _2 (s)=s$ (homogeneous field driving) and focus on effects of the $XX$ interactions. First consider the case of the $XX$ interaction between the clusters. Since $\hat{\bm{m}}_a$ is the sum of a large number of spins, it reduces to a classical variable in the thermodynamic limit $N\to\infty$, which significantly facilitates the analysis. As a consequence, we can calculate the magnetization $\bm{m}_a$ for each cluster in the ground state as detailed in Appendix~\ref{ssec.infmethod_cllimit}.

\begin{figure*}
\centering
\includegraphics{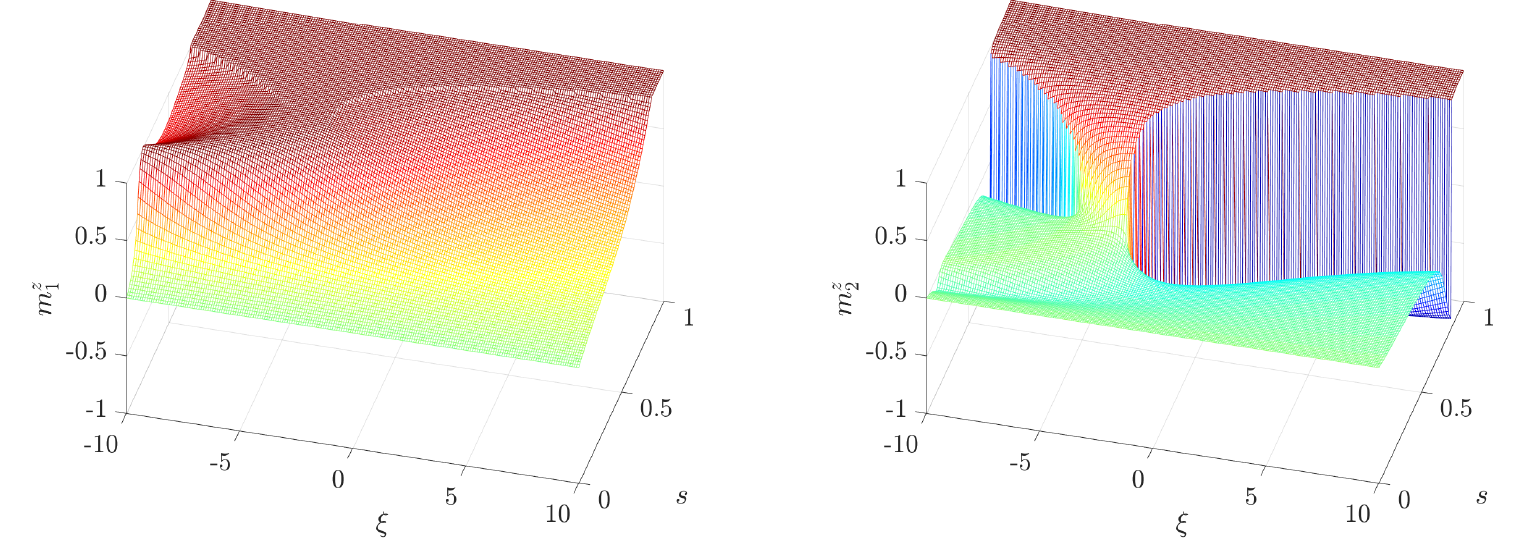}
\caption{(Color online) Magnetizations $m_1^z$ and $m_2^z$ of the weak-strong cluster problem with dense intercluster interactions for $\gamma _1 (s)=\gamma _2 (s)=s$ and $(\xi _{11} ,\xi _{22} ,\xi _{12})=(0,0,\xi )$.} \label{fig.wscl_mf_mzXXbt}
\end{figure*}

The result for the magnetizations $m_1^z$ and $m_2^z$ is shown as functions of $s$ and $\xi =\xi _{12}$ for $\xi _{11} =\xi _{22} =0$ in Fig.~\ref{fig.wscl_mf_mzXXbt}. A first-order transition exists in the stoquastic Hamiltonian with $\xi\geq 0$ including the case without the $XX$ interactions ($\xi =0$), where the magnetization in the strong cluster $m_1^z$ slightly jumps and that in the weak cluster $m_2^z$ jumps from a negative value to a positive value (a large-scale spin flip). On the other hand, there is no transition for $-5\lesssim\xi\lesssim -3$. This means that the non-stoquastic $XX$ interaction between the clusters with an appropriate strength removes the first-order transition, while too small or too large ones cannot. This result confirms the conclusion obtained by exact diagonalization of the finite-size systems\cite{Albash2019}.

\begin{figure*}
\centering
\includegraphics{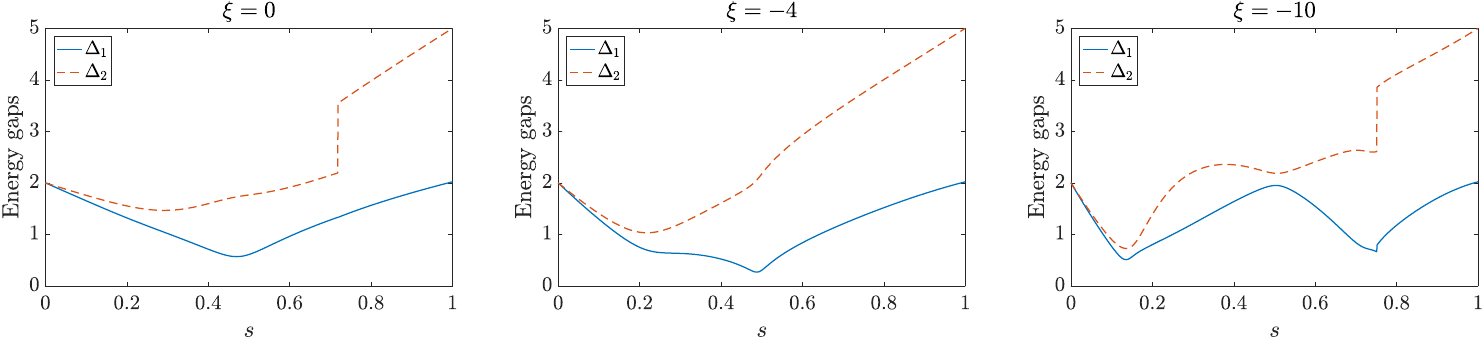}
\caption{(Color online) Two classes of the energy gap $\Delta _a$ created by the quasi-particle excitations $\hat{b}_a'^\dagger$ for the weak-strong cluster problem with dense intercluster interactions. We set $\gamma _1 (s)=\gamma _2 (s)=s$, $\xi _{11} =\xi _{22} =0$, and $\xi _{12} =\xi =0,-4,-10$.} \label{fig.wscl_mf_gapsXXbt}
\end{figure*}

We next calculate the energy gap between the ground state and the first excited state, which can be achieved by evaluating quantum fluctuations around the classical limit\cite{Seoane2012,Filippone2011} as detailed in Appendix~\ref{ssec.infmethod_qufluc}. We show the resulting energy gaps $\Delta _a~(a=1, 2)$ for $\xi _{11} =\xi _{22} =0$ and $\xi _{12} =\xi =0,-4,-10$ in Fig.~\ref{fig.wscl_mf_gapsXXbt}. Here, $\Delta _a$ denote the energy gaps created by the quasi-particle excitations $\hat{b}_a'^\dagger$ above the classical ground state. We calculated $\Delta _a$ by numerically diagonalizing the four-dimensional matrix $\mathcal{E}$ defined as Eq.~\eqref{eq.infmethod_2Admat} and multiplying the non-negative eigenvalues $\varepsilon _a$ by four (see Eq.~\eqref{eq.infmethod_elementarygaps}). The smaller gap $\Delta _1$ is equal to the energy gap between the ground and first excited states of the Hamiltonian $\hat{H} (s)$ except at the first-order transition point. The correct energy gap at a first-order transition is exponentially small as a function of the system size $N$\cite{Albash2019}, which cannot be evaluated by our method since our method gives the energy gap in the thermodynamic limit (see Appendix~\ref{ssec.infmethod_qufluc}). In general, the energy gaps $\Delta _a$ calculated by our method are discontinuous at first-order transitions due to the discontinuity of the magnetizations $\bm{m}_a$, although we cannot clearly see a discontinuous jump in the lower of the two gaps $\Delta _1$ for $\xi =0$ at least in our precision whereas the other $\Delta _2$ shows discontinuity. In the case of $\xi =-4$, the energy gaps $\Delta _a$ are continuous because there is no first-order transition.

\begin{figure}
\centering
\includegraphics{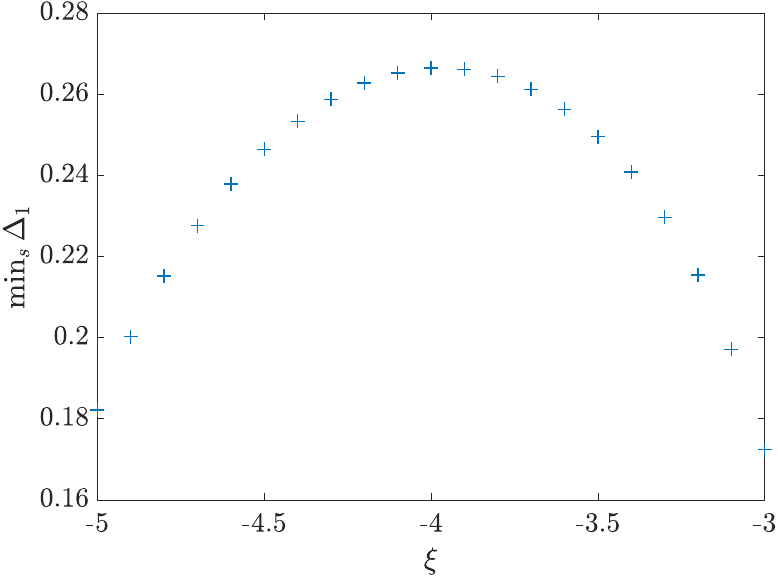}
\caption{(Color online) Minimum energy gap $\min _s \Delta _1$ of the weak-strong cluster problem with dense intercluster interactions for $\gamma _1 (s)=\gamma _2 (s)=s$, $\xi _{11} =\xi _{22} =0$, and $-5\leq\xi _{12} =\xi\leq -3$. In this region of $\xi$, there is no first-order transition and $\Delta _1$ is equal to the energy gap between the ground and first excited states of the Hamiltonian $\hat{H} (s)$ in the thermodynamic limit $N\to\infty$.} \label{fig.wscl_mf_mingapXXbt}
\end{figure}

Now we derive the minimum gap $\min _s \Delta _1$ for the range of $\xi$ where there is no first-order transition. The computation proceeds as in the previous calculation with details found in Appendix~\ref{ssec.infmethod_qufluc}. The result for $\xi _{11} =\xi _{22} =0$ and $-5\leq\xi _{12} =\xi\leq -3$ is shown in Fig.~\ref{fig.wscl_mf_mingapXXbt}. We can see that $\min _s \Delta _1$ is maximized at $\xi\approx -4.0$, which is consistent with the result in Ref.~\citen{Albash2019} (notice that $\lambda$ in Ref.~\citen{Albash2019} is equal to $-\xi /2$ in our definition).

\begin{figure*}
\centering
\includegraphics{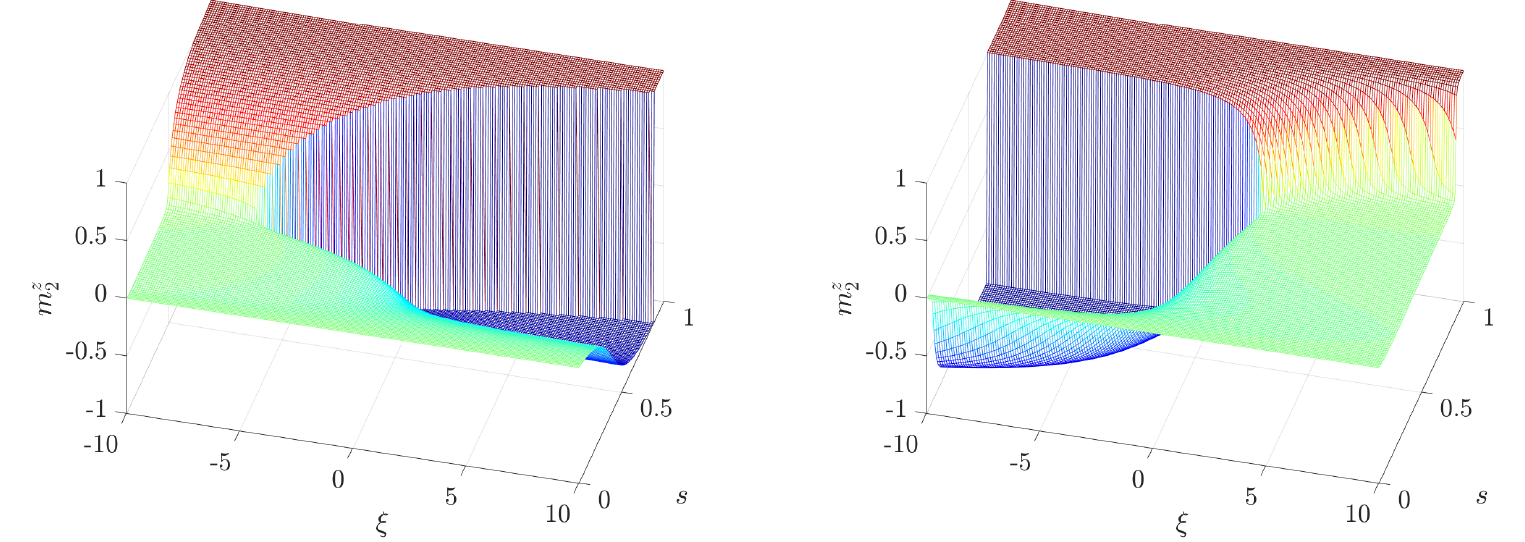}
\caption{(Color online) Magnetization in the weak cluster $m_2^z$ of the weak-strong cluster problem with dense intercluster interactions for $(\xi _{11} ,\xi _{22} ,\xi _{12})=(\xi ,0,0)$ (left) and for $(\xi _{11} ,\xi _{22} ,\xi _{12})=(0,\xi ,0)$ (right). We set $\gamma _1 (s)=\gamma _2 (s)=s$ in both cases.} \label{fig.wscl_mf_mzXXin}
\end{figure*}

Let us move on to the case of the $XX$ interaction in each cluster, which was not covered in Ref.~\citen{Albash2019}. We show the magnetization in the weak cluster $m_2^z$ for $(\xi _{11} ,\xi _{22} ,\xi _{12})=(\xi ,0,0),(0,\xi ,0)$ as functions of $s$ and $\xi$ in Fig.~\ref{fig.wscl_mf_mzXXin}. We find that the non-stoquastic $XX$ interaction in the strong cluster or the stoquastic $XX$ interaction in the weak cluster removes the first-order transition, while the other types of intracluster $XX$ interaction do not.

We can interpret these results as follows. In the case of the non-stoquastic $XX$ interaction in the strong cluster, $\lvert m_1^x \rvert$ becomes smaller and $m_1^z$ larger, which makes $m_2^z$ larger thanks to the ferromagnetic coupling between the clusters. In the case of the stoquastic $XX$ interaction in the weak cluster, $m_2^x$ becomes larger and $\lvert m_2^z \rvert$ smaller. Both of these types of $XX$ interaction prevent $m_2^z$ from being a large negative value due to the longitudinal field $h_2$, which reduces the possibility of a jump in $m_2^z$.

We also found that the first-order transition cannot be removed in the case where the $XX$ interaction is proportional to $\frac{1}{2} (\hat{m}_1^x +\hat{m}_2^x)^2$ (i.e., $(\xi _{11} ,\xi _{22} ,\xi _{12})=(\xi /2,\xi /2,\xi )$) regardless of the sign of the coefficient $\xi$, which is shown in Appendix~\ref{sec.totalcatalyst}. The result in the non-stoquastic case $\xi <0$ is in agreement with the numerical consequence given in Appendix~F of Ref.~\citen{Albash2019}.

\begin{figure*}
\centering
\includegraphics{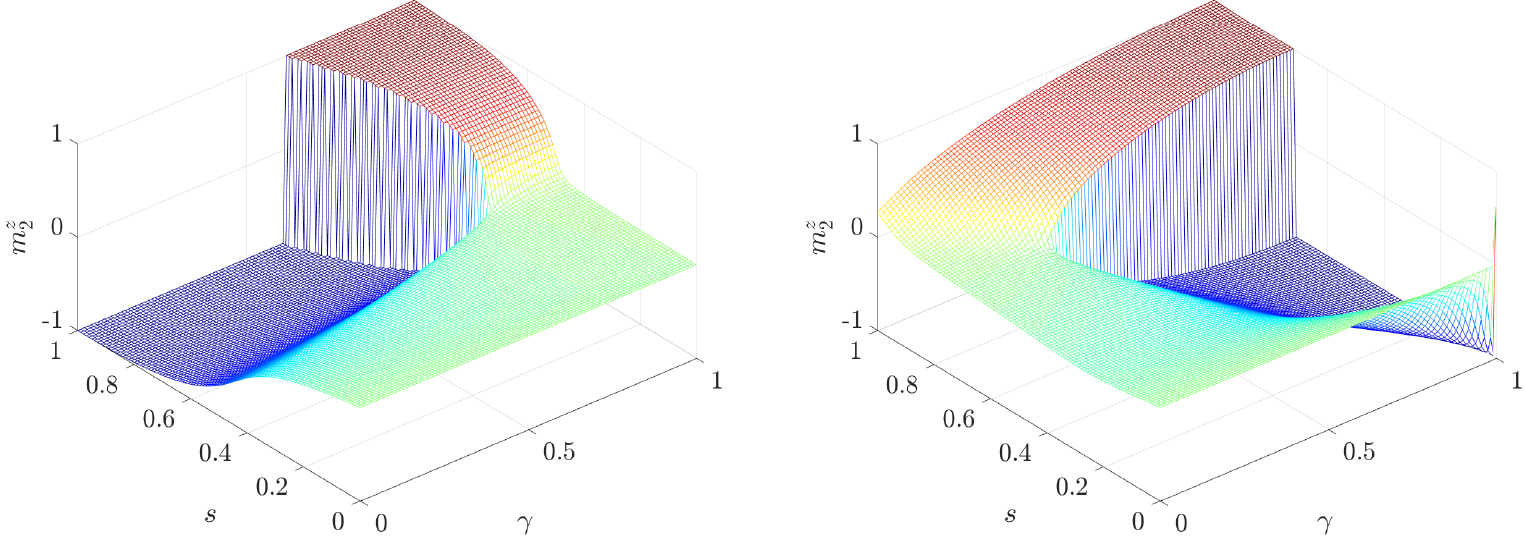}
\caption{(Color online) Magnetization in the weak cluster $m_2^z$ of the weak-strong cluster problem with dense intercluster interactions for $(\gamma _1 ,\gamma _2)=(\gamma ,s)$ (left) and for $(\gamma _1 ,\gamma _2)=(s,\gamma )$ (right). We set $\xi _{11} =\xi _{22} =\xi _{12} =0$ in both cases.} \label{fig.wscl_mf_mzihX}
\end{figure*}

We next consider the problem in which the transverse field is driven inhomogeneously and there is no $XX$ interaction. Then, the Hamiltonian is stoquastic. We show the magnetization in the weak cluster $m_2^z$ for $(\gamma _1 (s),\gamma _2 (s))=(\gamma (s),s),(s,\gamma (s))$ in Fig.~\ref{fig.wscl_mf_mzihX}, where the increasing function $\gamma (s)$ can be chosen arbitrarily as long as $\gamma (0)=0$ and $\gamma (1)=1$. Notice that the value of $m_2^z$ is indefinite at $s=0$ and $\gamma _2 =1$, where neither magnetic field nor interaction is applied to the weak cluster. We find that the weaker transverse field in the strong cluster and the stronger transverse field in the weak cluster can remove the first-order transition in the process of quantum annealing. The mechanism for removing the first-order transition is similar to the case of the non-stoquastic $XX$ interaction in the strong cluster or the stoquastic $XX$ interaction in the weak cluster.

\subsection{Weak-strong cluster problem with sparse interactions between clusters} \label{ssec.wscl_sp}

\begin{figure}
\centering
\includegraphics[scale=0.3]{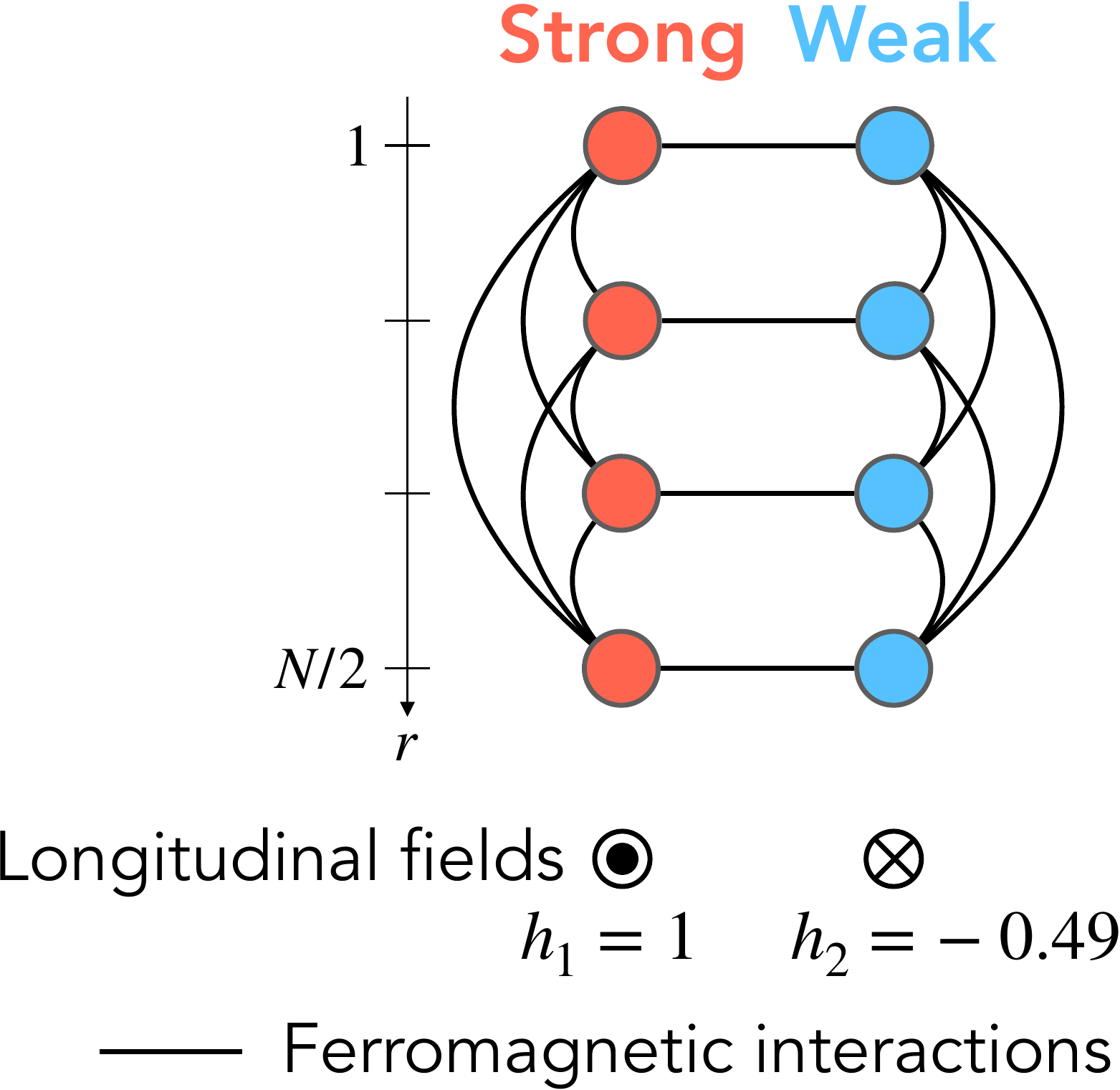}
\caption{(Color online) Weak-strong cluster problem with sparse interactions between clusters. The circles in the left (right) side denote the spins in the strong (weak) cluster.} \label{fig.wscl_sp_diagram}
\end{figure}

We now consider the weak-strong cluster problem whose interactions between the clusters are sparse. The problem Hamiltonian is defined as
\begin{align}
\hat{H}_\mathrm{p} & =-\sum _{r=1}^{N/2} (h_1 \hat{\sigma}_{1r}^z +h_2 \hat{\sigma}_{2r}^z) \notag \\
& \hphantom{{} = {}} -\frac{1}{N} \sum _{r,r'=1}^{N/2} (\hat{\sigma}_{1r}^z \hat{\sigma}_{1r'}^z +\hat{\sigma}_{2r}^z \hat{\sigma}_{2r'}^z)-\frac{1}{2} \sum _{r=1}^{N/2} \hat{\sigma}_{1r}^z \hat{\sigma}_{2r}^z ,
\end{align}
where the longitudinal field in the strong cluster is $h_1 =1$ and that in the weak cluster is $h_2 =-0.49$. Notice that the intercluster interactions exist only between the corresponding indices of the two clusters. We show the schematic diagram of the problem in Fig.~\ref{fig.wscl_sp_diagram}.

The quantum annealing Hamiltonian for this problem is given by
\begin{align}
\hat{H} (s) & =s\hat{H}_\mathrm{p} -(1-\gamma _1 (s))\sum _{r=1}^{N/2} \hat{\sigma}_{1r}^x -(1-\gamma _2(s))\sum _{r=1}^{N/2} \hat{\sigma}_{2r}^x \notag \\
& \hphantom{{} = {}} -s(1-s)\left(\frac{\xi _{11}}{N} \sum _{r,r'=1}^{N/2} \hat{\sigma}_{1r}^x \hat{\sigma}_{1r'}^x +\frac{\xi _{22}}{N} \sum _{r,r'=1}^{N/2} \hat{\sigma}_{2r}^x \hat{\sigma}_{2r'}^x \right. \notag \\
& \hphantom{{} =-s(1-s)} \left. {} +\frac{\xi _{12}}{2} \sum _{r=1}^{N/2} \hat{\sigma}_{1r}^x \hat{\sigma}_{2r}^x \right) , \label{eq.wscl_sp_hamiltonian}
\end{align}
where $s\in [0,1]$ is the dimensionless time. After taking the thermodynamic limit $N\to\infty$ and the zero-temperature limit, we calculate the magnetizations in the two clusters $m_1^z$ and $m_2^z$ using the imaginary-time path-integral formulation of the partition function and the saddle-point method with the static ansatz as detailed in Appendix~\ref{sec.sinfmethod}.

\begin{figure*}
\centering
\includegraphics{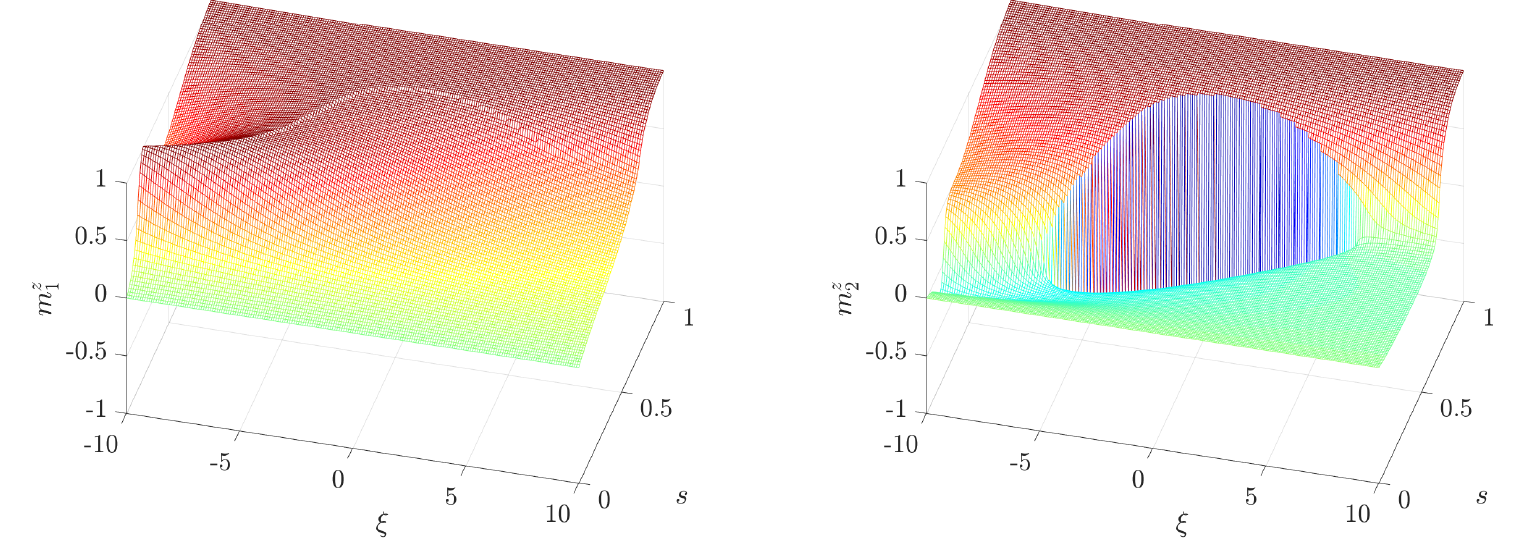}
\caption{(Color online) Magnetizations $m_1^z$ and $m_2^z$ of the weak-strong cluster problem with sparse intercluster interactions for $\gamma _1 (s)=\gamma _2 (s)=s$ and $(\xi _{11} ,\xi _{22} ,\xi _{12})=(0,0,\xi )$.} \label{fig.wscl_sp_mzXXbt}
\end{figure*}

We show the magnetizations $m_1^z$ and $m_2^z$ for $\gamma _1 (s)=\gamma _2 (s)=s$ and $(\xi _{11} ,\xi _{22} ,\xi _{12})=(0,0,\xi )$ in Fig.~\ref{fig.wscl_sp_mzXXbt}. We find that while the uniform transverse-field driver $\xi =0$ causes a first-order transition, both of the non-stoquastic $XX$ interaction between the clusters $\xi <0$ and the stoquastic one $\xi >0$ can remove the transition. In contrast to the case of dense interactions discussed in Section~\ref{ssec.wscl_mf}, there is no transition for large positive $\xi$ and too large negative $\xi$.

\begin{figure*}
\centering
\includegraphics{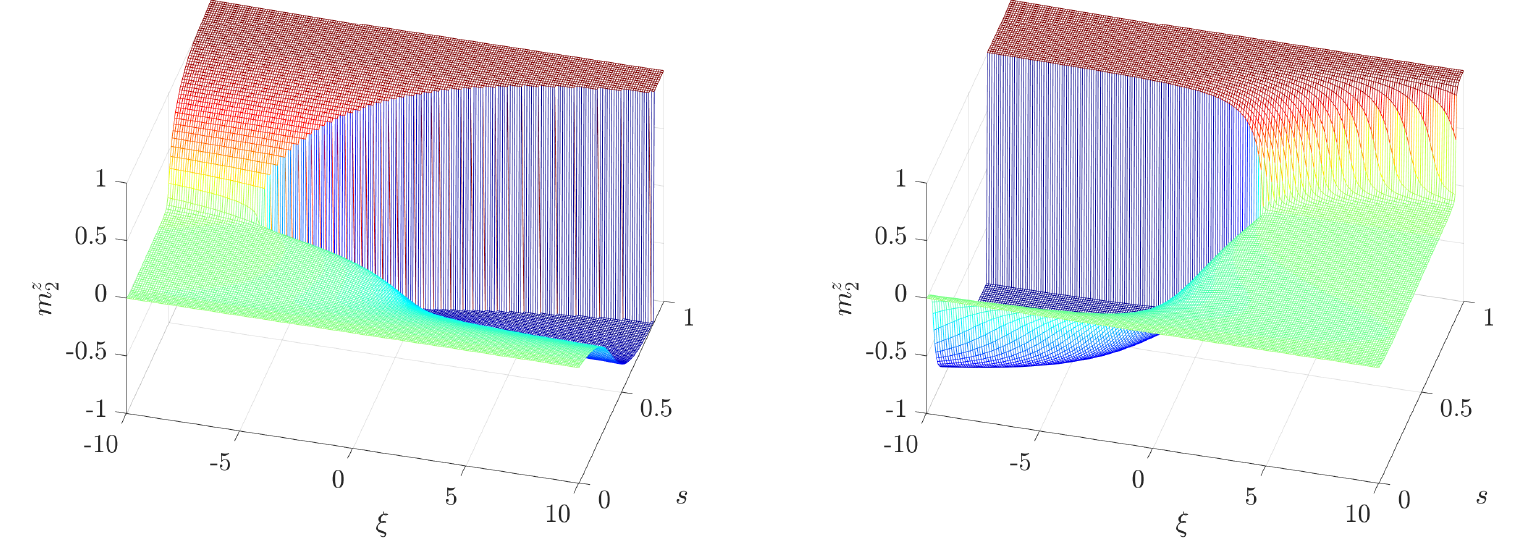}
\caption{(Color online) Magnetization in the weak cluster $m_2^z$ of the weak-strong cluster problem with sparse intercluster interactions for $(\xi _{11} ,\xi _{22} ,\xi _{12})=(\xi ,0,0)$ (left) and for $(\xi _{11} ,\xi _{22} ,\xi _{12})=(0,\xi ,0)$ (right). We set $\gamma _1 (s)=\gamma _2 (s)=s$ in both cases.} \label{fig.wscl_sp_mzXXin}
\end{figure*}

\begin{figure*}
\centering
\includegraphics{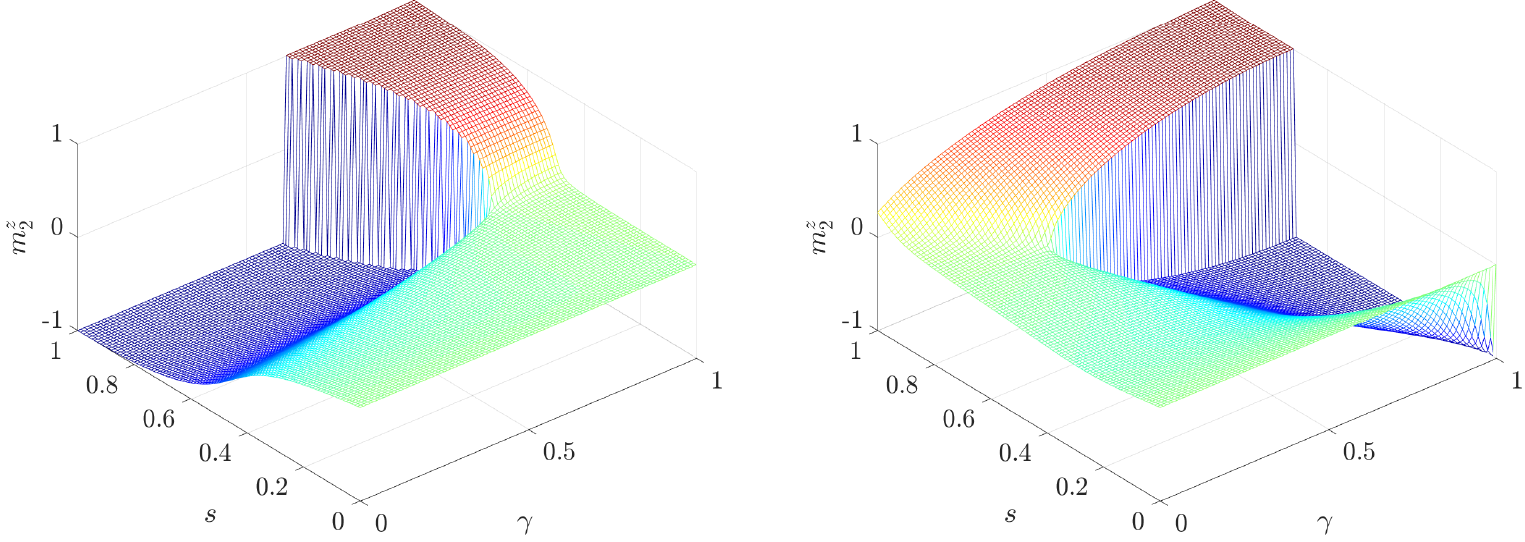}
\caption{(Color online) Magnetization in the weak cluster $m_2^z$ of the weak-strong cluster problem with sparse intercluster interactions for $(\gamma _1 ,\gamma _2)=(\gamma ,s)$ (left) and $(\gamma _1 ,\gamma _2)=(s,\gamma )$ (right). We set $\xi _{11} =\xi _{22} =\xi _{12} =0$ in both cases.} \label{fig.wscl_sp_mzihX}
\end{figure*}

On the other hand, Fig.~\ref{fig.wscl_sp_mzXXin} shows that the behavior of the magnetization $m_2^z$ in the case of the $XX$ interaction in each cluster, $\gamma _1 (s)=\gamma _2 (s)=s$ and $(\xi _{11} ,\xi _{22} ,\xi _{12})=(\xi ,0,0),(0,\xi ,0)$, is similar to that for the problem with dense intercluster interactions. In addition, the behavior of the magnetization $m_2^z$ under inhomogeneous driving of the transverse field (i.e., $(\gamma _1 (s),\gamma _2 (s))=(\gamma (s),s),(s,\gamma (s))$ and $\xi _{11} =\xi _{22} =\xi _{12} =0$) resembles that in the case of dense intercluster interactions as can be seen in Fig.~\ref{fig.wscl_sp_mzihX}.

Before concluding, we notice that the first-order transition is unavoidable in the case of the total $XX$ interaction, $\gamma _1 (s)=\gamma _2 (s)=s$ and $(\xi _{11} ,\xi _{22} ,\xi _{12})=(\xi /2,\xi /2,\xi )$, as shown in Appendix~\ref{sec.totalcatalyst}.

\section{Conclusion} \label{sec.conc}

We have studied the phase transitions of two weak-strong cluster problems with the ultimate goal to reveal what types of catalyst remove troublesome first-order transitions in quantum annealing. The Hamiltonian of each model consists of longitudinal fields and ferromagnetic $ZZ$ interactions in and between the clusters as well as transverse fields and $XX$ interactions (catalysts). The longitudinal fields in the weak and strong clusters have opposite directions and different strengths, which causes a first-order phase transition in the absence of the catalysts. The difference between the two problems is the connectivity between the clusters: One has dense (all-to-all) interactions between the clusters and the other has sparse interactions.

We solved the problem by a semi-classical method and found that stoquastic or non-stoquastic catalysts can remove the first-order transition for the model with all-to-all interactions between the clusters. More precisely, we first showed that the transition disappears if a non-stoquastic catalyst is appended between the clusters with an appropriate strength while the transition persists if the catalyst is stoquastic, which is consistent with the already known result of numerical diagonalization of finite-size systems\cite{Albash2019}. We also calculated the energy gap in the thermodynamic limit analytically and identified the optimal strength of the non-stoquastic $XX$ interaction between the clusters that maximizes the minimum energy gap. The result again confirms the consequence of the numerical study\cite{Albash2019}. In addition to the non-stoquastic catalyst between the clusters, we found other protocols to eliminate the first-order transition, namely, a non-stoquastic catalyst in the strong cluster, a stoquastic catalyst in the weak cluster, and inhomogeneous transverse-field driving in which the transverse field is weaker in the strong cluster or stronger in the weak cluster. The latter result confirms general observations in previous studies on the usefulness of inhomogeneous field driving\cite{Susa2018,Susa2018b,Adame2018,Hartmann2019}.

We next analyzed the problem with sparse interactions between the clusters by evaluating the partition function in the thermodynamic limit and the zero-temperature limit. Then, we found generally similar results as in the case with all-to-all interactions, except that a stoquastic catalyst between the clusters as well as a non-stoquastic one can remove the first-order transition.

It is noteworthy that our results are rare examples of two-body interacting systems for which the removal of first-order transitions with stoquastic or non-stoquastic catalysts has been shown analytically. Although it is generally difficult to predict for a given real-world optimization problem which type of catalyst (stoquastic or non-stoquastic) or inhomogeneous driving is effective to enhance the performance of quantum annealing, it is likely to be useful to introduce many-body drivers ($XX$ interactions with adjustable sign and strength) and inhomogeneous transverse-field driving in the design of hardware of quantum annealing. To better understand the effects of stoquastic and non-stoquastic catalysts and inhomogeneity in the transverse field, analytical and numerical studies of many other problems are highly desired, especially in the cases with sparse connectivity to represent realistic situations.

\section*{Acknowledgment}

We thank Tameem Albash for useful comments. The work of KT is supported by JSPS KAKENHI Grant No.~17J09218 and that of HN is by JSPS KAKENHI Grant No.~26287086. The work of HN is financially supported also by the Office of the Director of National Intelligence (ODNI), Intelligence Advanced Research Projects Activity (IARPA), via U.S. Army Research Office Contract No.~W911NF-17-C-0050. The views and conclusions contained herein are those of the authors and should not be interpreted as necessarily representing the official policies or endorsements, either expressed or implied, of the ODNI, IARPA, or the U.S. Government. The U.S. Government is authorized to reproduce and distribute reprints for Governmental purposes notwithstanding any copyright annotation thereon.

\appendix

\section{Analysis of an Infinite-Range System Consisting of Several Subsystems} \label{sec.infmethod}

We analyze a mean-field spin system consisting of several subsystems by use of the semi-classical method. The system is supposed to have infinite-range (all-to-all) interactions in each subsystem and between subsystems. We first take the classical limit to calculate the magnetization and next include quantum fluctuations to evaluate the energy gap.

\subsection{Classical limit} \label{ssec.infmethod_cllimit}

Let us consider a spin system which consists of several subsystems. Let $N$ be the total number of spins and $A=\mathcal{O} (N^0)$ be the number of subsystems. The problem in the main text has $A=2$ but we develop a general argument here for possible future convenience. Suppose that each subsystem has the equal number of spins. We denote the Pauli operator at site $(a,r)$ by $\hat{\bm{\sigma}}_{ar} =(\hat{\sigma}_{ar}^\alpha )_{\alpha =x,y,z}$, where $a=1,\dots ,A$ is the subsystem index and $r=1,\dots ,N/A$ is the site index in each subsystem.

We consider the Hamiltonian $\hat{H}$ which is written as a function of the total spin operators for the subsystems $\hat{\bm{S}}_a =\frac{1}{2} \sum _r \hat{\bm{\sigma}}_{ar}$. The operators $\hat{\bm{S}}_a$ satisfy the commutation relations $[\hat{S}_a^\alpha ,\hat{S}_b^\beta ]=i\delta _{ab} \sum _\gamma \epsilon ^{\alpha\beta\gamma} \hat{S}_a^\gamma$, where $\delta _{ab}$ is the Kronecker delta and $\epsilon ^{\alpha\beta\gamma}$ is the Levi-Civita symbol. Since $[\hat{H} ,\hat{\bm{S}}_a^2]=0$ and the initial state of quantum annealing is the state in which all the spins point in the $x$-direction, the time evolution of quantum annealing occurs in the eigenspace of $\hat{\bm{S}}_a^2$ with $S_a =\frac{N}{2A} =:S$, where $S_a (S_a +1)$ are the eigenvalues of $\hat{\bm{S}}_a^2$. Therefore, we can consider that $\hat{\bm{S}}_a$ are spin-($S=\frac{N}{2A}$) operators and the system consists of $A$ interacting large spins.

Defining the magnetization operator for each subsystem as
\begin{equation}
\hat{\bm{m}}_a =(\hat{m}_a^\alpha )_{\alpha =x,y,z} =\frac{A}{N} \sum _r \hat{\bm{\sigma}}_{ar} =\frac{\hat{\bm{S}}_a}{S} ,
\end{equation}
we can write the Hamiltonian $\hat{H}$ as $\hat{H} =Nh(\{\hat{\bm{m}}_a \} )$. We assume that $h(\{\hat{\bm{m}}_a \})$ is a polynomial of degree $P=\mathcal{O} (N^0)$, i.e. a linear combination of $\hat{m}_{a_1}^{\alpha _1} \dotsm\hat{m}_{a_p}^{\alpha _p}$ ($p=0,1,\dots ,P$), and the coefficients are of $\mathcal{O} (N^0)$. Now we take the thermodynamic limit $N\to\infty\iff S\to\infty$. In this limit, the non-commutativity of the components of $\hat{\bm{m}}_a$ is negligible and $\hat{\bm{m}}_a^2$ approaches unity:
\begin{gather}
[\hat{m}_a^\alpha ,\hat{m}_b^\beta ]=\frac{1}{S} i\delta _{ab} \sum _\gamma \epsilon ^{\alpha\beta\gamma} \hat{m}_a^\gamma \to 0, \\
\hat{\bm{m}}_a^2 =\frac{\hat{\bm{S}}_a^2}{S^2} =\frac{S(S+1)}{S^2} \to 1.
\end{gather}
These equations mean that we can regard the operators $\hat{\bm{m}}_a$ as classical unit vectors $\bm{m}_a$ in the limit $S\to\infty$. Since the eigenvalues of the operator $\hat{m}_a^\alpha$ are $-1,-1+1/S,\dots ,+1$, each component $m_a^\alpha$ of the vector $\bm{m}_a$ takes continuous values in $[-1,1]$.

Accordingly, the ground state of $\hat{H}$ in the limit $S\to\infty$ is given by the vectors $\bm{m}_a$ which minimize the energy density $h(\{\bm{m}_a \} )$ subject to the constraints $\bm{m}_a^2 =1$. The partial derivatives of the function $h(\{\bm{m}_a \})+\sum _a \frac{\mu _a}{2} (\bm{m}_a^2 -1)$ vanish at the minimum point, where $\mu _a$ are the Lagrange multipliers:
\begin{equation}
\frac{\partial h}{\partial\bm{m}_a} =-\mu _a \bm{m}_a . \label{eq.infmethod_minenergy}
\end{equation}

\subsection{Quantum fluctuation} \label{ssec.infmethod_qufluc}

In order to derive the energy gap, we extend the method used in Refs.~\citen{Seoane2012} and \citen{Filippone2011} to the system of several large spins. First we wish to expand the spin operators $\hat{\bm{S}}_a$ around the classical limit. We introduce rotated spin operators $\hat{\bm{S}}_a'$ whose $z$-components are the spin operators in the directions of $\bm{m}_a =:\bm{e}_a'^z$. We choose unit vectors $\bm{e}_a'^x$ and $\bm{e}_a'^y$ such that $\{\bm{e}_a'^\alpha \} _{\alpha =x,y,z}$ is an orthonormal set and $\bm{e}_a'^x \times\bm{e}_a'^y =\bm{e}_a'^z$. Defining the components of $\hat{\bm{S}}_a'$ as $\hat{S}_a'^\alpha =\bm{e}_a'^\alpha \cdot\hat{\bm{S}}_a$, we obtain
\begin{equation}
\hat{\bm{S}}_a' =T_a^\mathrm{T} \hat{\bm{S}}_a \iff\hat{\bm{S}}_a =T_a \hat{\bm{S}}_a',
\end{equation}
where $T_a =(T_a^{\alpha\beta})_{\alpha ,\beta =x,y,z} =(\bm{e}_a'^x ,\bm{e}_a'^y ,\bm{e}_a'^z)\in\mathrm{SO} (3)$ is a special orthogonal matrix. Then, $\hat{\bm{S}}_a'$ satisfy the commutation relations $[\hat{S}_a'^\alpha ,\hat{S}_b'^\beta ]=i\delta _{ab} \sum _\gamma \epsilon ^{\alpha\beta\gamma} \hat{S}_a'^\gamma$.

Let $\hat{\bm{m}}_a'=\hat{\bm{S}}_a'/S$ and $\hat{m}_a'^\pm =\hat{m}_a'^x \pm i\hat{m}_a'^y$. We perform the Holstein-Primakoff transformation\cite{Holstein1940}
\begin{align}
\hat{m}_a'^+ & =\sqrt{\frac{2}{S}} \sqrt{1-\frac{1}{2S} \hat{b}_a^\dagger \hat{b}_a} \,\hat{b}_a , \\
\hat{m}_a'^- & =\sqrt{\frac{2}{S}} \,\hat{b}_a^\dagger \sqrt{1-\frac{1}{2S} \hat{b}_a^\dagger \hat{b}_a} , \\
\hat{m}_a'^z & =1-\frac{1}{S} \hat{b}_a^\dagger \hat{b}_a ,
\end{align}
where $\hat{b}_a$ are bosonic operators satisfying $[\hat{b}_a ,\hat{b}_b^\dagger ]=\delta _{ab}$ and $[\hat{b}_a ,\hat{b}_b]=0$.

The fact that $\hat{m}_a'^z =\bm{m}_a \cdot\hat{\bm{m}}_a$ approaches unity in the classical limit $S\to\infty$ implies that the number operators $\hat{n}_a =\hat{b}_a^\dagger \hat{b}_a$ take values sufficiently smaller than $S$ in the low-energy states for large $S$. Expanding the operators in $S^{-1}$ results in
\begin{align}
\hat{m}_a'^+ & =\sqrt{\frac{2}{S}} \,\hat{b}_a +\mathcal{O} (S^{-3/2}), \\
\hat{m}_a'^- & =\sqrt{\frac{2}{S}} \,\hat{b}_a^\dagger +\mathcal{O} (S^{-3/2}).
\end{align}
We thus obtain
\begin{align}
\hat{m}_a'^x & =\frac{1}{\sqrt{S}} \hat{q}_a +\mathcal{O} (S^{-3/2}), \\
\hat{m}_a'^y & =\frac{1}{\sqrt{S}} \hat{p}_a +\mathcal{O} (S^{-3/2}), \\
\hat{m}_a'^z & =1-\frac{1}{S} \hat{n}_a ,
\end{align}
where $\hat{q}_a =(\hat{b}_a +\hat{b}_a^\dagger )/\sqrt{2}$ and $\hat{p}_a =(\hat{b}_a -\hat{b}_a^\dagger )/(\sqrt{2} \, i)$ are the coordinate and momentum operators for the $a$th harmonic oscillator, respectively. The operators $\hat{q}_a$ and $\hat{p}_a$ satisfy the canonical commutation relations $[\hat{q}_a ,\hat{p}_b]=i\delta _{ab}$.

Then, we find that the original magnetization operators $\hat{\bm{m}}_a =T_a \hat{\bm{m}}_a'$ are expanded as
\begin{equation}
\hat{\bm{m}}_a =\bm{m}_a +\frac{1}{\sqrt{S}} (\bm{e}_a'^x \hat{q}_a +\bm{e}_a'^y \hat{p}_a)-\frac{1}{S} \bm{m}_a \hat{n}_a +\mathcal{O} (S^{-3/2}).
\end{equation}
The Hamiltonian density operator has the expansion
\begin{align}
h(\{\hat{\bm{m}}_a \} ) & =h(\{\bm{m}_a \} ) \notag \\
& \hphantom{{} = {}} +\sum _a \frac{\partial h}{\partial\bm{m}_a} \cdot\left(\frac{1}{\sqrt{S}} (\bm{e}_a'^x \hat{q}_a +\bm{e}_a'^y \hat{p}_a)-\frac{1}{S} \bm{m}_a \hat{n}_a \right) \notag \\
& \hphantom{{} = {}} +\frac{1}{2S} \sum _{ab} (\bm{e}_a'^x \hat{q}_a +\bm{e}_a'^y \hat{p}_a)^\mathrm{T} \frac{\partial ^2 h}{\partial\bm{m}_a \,\partial\bm{m}_b^\mathrm{T}} (\bm{e}_b'^x \hat{q}_b +\bm{e}_b'^y \hat{p}_b) \notag \\
& \hphantom{{} = {}} +\frac{1}{S} c+\mathcal{O} (S^{-3/2}),
\end{align}
where we defined the Hessian matrix as $\frac{\partial ^2 h}{\partial\bm{m}_a \,\partial\bm{m}_b^\mathrm{T}} :=\left(\frac{\partial}{\partial\bm{m}_a} \right)\left(\frac{\partial}{\partial\bm{m}_b} \right) ^\mathrm{T} h$. In the above equation, the $c$-number $c$ arises from the non-commutativity of $\bm{e}_a'^x \hat{q}_a +\bm{e}_a'^y \hat{p}_a$ and $\bm{e}_b'^x \hat{q}_b +\bm{e}_b'^y \hat{p}_b$:
\begin{align}
& \hphantom{{} = {}} [(\bm{e}_a'^x \hat{q}_a +\bm{e}_a'^y \hat{p}_a)^\alpha ,(\bm{e}_b'^x \hat{q}_b +\bm{e}_b'^y \hat{p}_b)^\beta ] \notag \\
& =i\delta _{ab} ((\bm{e}_a'^x)^\alpha (\bm{e}_a'^y)^\beta -(\bm{e}_a'^y)^\alpha (\bm{e}_a'^x)^\beta ) \notag \\
& =i\delta _{ab} (T_a^{\alpha x} T_a^{\beta y} -T_a^{\alpha y} T_a^{\beta x}).
\end{align}
However, the value of $c$ is not needed for determining the energy gap.

Since the magnetizations $\bm{m}_a$ in the ground state satisfy Eq.~\eqref{eq.infmethod_minenergy} and $\bm{e}_a'^x$ and $\bm{e}_a'^y$ are orthogonal to $\bm{m}_a$, we find the expansion of the Hamiltonian density operator
\begin{equation}
h(\{\hat{\bm{m}}_a \} )=h(\{\bm{m}_a \} )+\frac{1}{S} \hat{\varepsilon} +\mathcal{O} (S^{-3/2}),
\end{equation}
where
\begin{equation}
\hat{\varepsilon} =\sum _a \mu _a \hat{n}_a +\frac{1}{2} \sum _{ab} [h_{ab}^{xx} \hat{q}_a \hat{q}_b +h_{ab}^{yy} \hat{p}_a \hat{p}_b +h_{ab}^{xy} (\hat{q}_a \hat{p}_b +\hat{p}_b \hat{q}_a)]+c \label{eq.infmethod_undiaggapop}
\end{equation}
and
\begin{align}
\mu _a & =-\frac{\partial h}{\partial\bm{m}_a} \cdot\bm{m}_a , \\
h_{ab}^{\alpha\beta} & :=(\bm{e}_a'^\alpha )^\mathrm{T} \frac{\partial ^2 h}{\partial\bm{m}_a \,\partial\bm{m}_b^\mathrm{T}} \bm{e}_b'^\beta .
\end{align}
For $\alpha =\beta$, $h_{ab}^{\alpha\alpha}$ is symmetric under the exchange of the lower indices: $h_{ab}^{\alpha\alpha} =h_{ba}^{\alpha\alpha}$.

Let us diagonalize the operator $\hat{\varepsilon}$. We perform the Bogoliubov transformation
\begin{align}
\hat{b}_a' & =\sum _b (U_{ab} \hat{b}_b +V_{ab} \hat{b}_b^\dagger ), \label{eq.infmethod_bogoliubov1} \\
\hat{b}_a'^\dagger & =\sum _b (V_{ab}^* \hat{b}_b +U_{ab}^* \hat{b}_b^\dagger ) \label{eq.infmethod_bogoliubov2}
\end{align}
and assume that the new bosonic operators $\hat{b}_a'$ diagonalize $\hat{\varepsilon}$:
\begin{equation}
\hat{\varepsilon} =\sum _a \varepsilon _a \hat{b}_a'^\dagger \hat{b}_a'+c'. \label{eq.infmethod_diaggapop}
\end{equation}
Here, $\varepsilon _a$ and $c'$ should be real numbers for $\hat{\varepsilon}$ to be Hermitian and $\hat{b}_a'$ should satisfy the commutation relations
\begin{equation}
[\hat{b}_a',\hat{b}_b'^\dagger ]=\delta _{ab} ,\quad [\hat{b}_a',\hat{b}_b']=0. \label{eq.infmethod_commnewbosons}
\end{equation}

It follows from Eq.~\eqref{eq.infmethod_undiaggapop} that
\begin{align}
[\hat{q}_a ,\hat{\varepsilon}] & =i\sum _b [(\mu _a \delta _{ab} +h_{ab}^{yy})\hat{p}_b +h_{ba}^{xy} \hat{q}_b] \notag \\
& =\frac{1}{\sqrt{2}} \sum _b [(\mu _a \delta _{ab} +h_{ab}^{yy} +ih_{ba}^{xy})\hat{b}_b \notag \\
& \hphantom{{} =\frac{1}{\sqrt{2}} \sum _b [} -(\mu _a \delta _{ab} +h_{ab}^{yy} -ih_{ba}^{xy})\hat{b}_b^\dagger ], \\
[\hat{p}_a ,\hat{\varepsilon}] & =-i\sum _b [(\mu _a \delta _{ab} +h_{ab}^{xx})\hat{q}_b +h_{ab}^{xy} \hat{p}_b] \notag \\
& =\frac{1}{\sqrt{2} \, i} \sum _b [(\mu _a \delta _{ab} +h_{ab}^{xx} -ih_{ab}^{xy})\hat{b}_b \notag \\
& \hphantom{{} =\frac{1}{\sqrt{2} \, i} \sum _b [} +(\mu _a \delta _{ab} +h_{ab}^{xx} +ih_{ab}^{xy})\hat{b}_b^\dagger ].
\end{align}
Combining these equations with Eq.~\eqref{eq.infmethod_bogoliubov1}, we derive
\begin{align}
[\hat{b}_a',\hat{\varepsilon}] & =\sum _b (U_{ab} [\hat{b}_b ,\hat{\varepsilon}]+V_{ab} [\hat{b}_b^\dagger ,\hat{\varepsilon}]) \notag \\
& =\frac{1}{\sqrt{2}} \sum _b \{ (U_{ab} +V_{ab})[\hat{q}_b ,\hat{\varepsilon}]+i(U_{ab} -V_{ab})[\hat{p}_b ,\hat{\varepsilon}]\} \notag \\
& =\sum _c \sum _b \{ [U_{ab} (\mu _b \delta _{bc} +Z_{bc}^+)-V_{ab} Z_{bc}^-]\hat{b}_c \notag \\
& \hphantom{{} =\sum _c \sum _b \{} +[U_{ab} Z_{bc}^{-*} -V_{ab} (\mu _b \delta _{bc} +Z_{bc}^{+*})]\hat{b}_c^\dagger \}, \label{eq.infmethod_commnewbosongap1}
\end{align}
where
\begin{equation}
Z_{ab}^\pm :=\frac{h_{ab}^{xx} \pm h_{ab}^{yy} -i(h_{ab}^{xy} \mp h_{ba}^{xy})}{2}
\end{equation}
and $Z_{ab}^{\pm *} :=(Z_{ab}^\pm )^*$. On the other hand, Eq.~\eqref{eq.infmethod_diaggapop} yields the commutation relation
\begin{equation}
[\hat{b}_a',\hat{\varepsilon}]=\varepsilon _a \hat{b}_a'=\varepsilon _a \sum _c (U_{ac} \hat{b}_c +V_{ac} \hat{b}_c^\dagger ). \label{eq.infmethod_commnewbosongap2}
\end{equation}
Comparing the coefficients of $\hat{b}_c$ and $\hat{b}_c^\dagger$ in Eqs.~\eqref{eq.infmethod_commnewbosongap1} and \eqref{eq.infmethod_commnewbosongap2} results in
\begin{align}
\varepsilon _a U_{ac} & =\sum _b [U_{ab} (\mu _b \delta _{bc} +Z_{bc}^+)-V_{ab} Z_{bc}^-], \label{eq.infmethod_eigeqcomp1} \\
\varepsilon _a V_{ac} & =\sum _b [U_{ab} Z_{bc}^{-*} -V_{ab} (\mu _b \delta _{bc} +Z_{bc}^{+*})]. \label{eq.infmethod_eigeqcomp2}
\end{align}
Notice that we can derive the equivalent equations by calculating $[\hat{b}_a'^\dagger ,\hat{\varepsilon}]$ with Eqs.~\eqref{eq.infmethod_undiaggapop}, \eqref{eq.infmethod_bogoliubov2}, and \eqref{eq.infmethod_diaggapop}.

Let us define the $A$-dimensional matrices
\begin{equation}
M=(\mu _a \delta _{ab}),\quad Z^\pm =(Z_{ab}^\pm)
\end{equation}
and the $A$-dimensional vectors
\begin{equation}
\bm{u}_a =(U_{a1} ,\dots ,U_{aA})^\mathrm{T} ,\quad\bm{v}_a =(V_{a1} ,\dots ,V_{aA})^\mathrm{T} .
\end{equation}
Then, the set of Eqs.~\eqref{eq.infmethod_eigeqcomp1} and \eqref{eq.infmethod_eigeqcomp2} is written as the eigenvalue equation
\begin{equation}
\bm{\psi}_a^\mathrm{T} \mathcal{E} =\bm{\psi}_a^\mathrm{T} \varepsilon _a , \label{eq.infmethod_eigeq}
\end{equation}
where
\begin{equation}
\mathcal{E} :=
\begin{pmatrix}
M+Z^+ & Z^{-*} \\
-Z^- & -M-Z^{+*}
\end{pmatrix}
\label{eq.infmethod_2Admat}
\end{equation}
is a $2A$-dimensional matrix and $\bm{\psi}_a :=(\bm{u}_a^\mathrm{T} ,\bm{v}_a^\mathrm{T})^\mathrm{T}$ is a $2A$-dimensional vector. Equation~\eqref{eq.infmethod_eigeq} shows that $\varepsilon _a$ are the eigenvalues of $\mathcal{E}$ and $\bm{\psi}_a^\mathrm{T}$ are the corresponding left eigenvectors.

We can show that $\varepsilon _a \in\mathbb{R}$ holds if $\lvert\bm{u}_a \rvert ^2 \not=\lvert\bm{v}_a \rvert ^2$. In the following, we assume that all of the eigenvalues of $\mathcal{E}$ are real numbers. Notice that the eigenvalue equation~\eqref{eq.infmethod_eigeq} yields another eigenvalue equation
\begin{equation}
\bm{\psi}_{-a}^\mathrm{T} \mathcal{E} =\bm{\psi}_{-a}^\mathrm{T} \varepsilon _{-a} ,
\end{equation}
where
\begin{equation}
\bm{\psi}_{-a} :=(\bm{u}_{-a}^\mathrm{T} ,\bm{v}_{-a}^\mathrm{T})^\mathrm{T} :=(\bm{v}_a^\dagger ,\bm{u}_a^\dagger )^\mathrm{T} ,\quad\varepsilon _{-a} :=-\varepsilon _a .
\end{equation}
This means that the $2A$-dimensional matrix $\mathcal{E}$ has $2A$ eigenvalues $\varepsilon _{\pm 1} ,\dots ,\varepsilon _{\pm A}$. Let $\varepsilon _1 ,\dots ,\varepsilon _A$ be non-negative eigenvalues of $\mathcal{E}$, which become the frequencies of the quasi-particles created by $\hat{b}_a'^\dagger$.

The commutation relations~\eqref{eq.infmethod_commnewbosons} are equivalent to the following constraints on $\bm{u}_a$ and $\bm{v}_a$:
\begin{equation}
\bm{u}_a^\mathrm{T} \bm{u}_b^* -\bm{v}_a^\mathrm{T} \bm{v}_b^* =\delta _{ab} ,\quad\bm{u}_a^\mathrm{T} \bm{v}_b -\bm{v}_a^\mathrm{T} \bm{u}_b =0\quad (a,b=1,\dots ,A).
\end{equation}
These constraints are rewritten as the pseudo orthonormality of $\bm{\psi}_a$,
\begin{equation}
\bm{u}_a^\mathrm{T} \bm{u}_b^* -\bm{v}_a^\mathrm{T} \bm{v}_b^* =\sgn (a)\delta _{ab} \quad (a,b=\pm 1,\dots ,\pm A). \label{eq.infmethod_pseudoorthnormeigvecs}
\end{equation}
We can show that Eq.~\eqref{eq.infmethod_eigeq} automatically yields Eq.~\eqref{eq.infmethod_pseudoorthnormeigvecs} if $\varepsilon _a \not=\varepsilon _b$. For each degenerate eigenvalue, we impose the constraint~\eqref{eq.infmethod_pseudoorthnormeigvecs} on the corresponding eigenvectors. The constraint for $a=b$, $\lvert\bm{u}_a \rvert ^2 -\lvert\bm{v}_a \rvert ^2 =\sgn (a)$, gives the normalization condition of the eigenvectors $\bm{\psi}_a$.

When $\varepsilon _a$ are real numbers for all $a$ and the pseudo orthonormality~\eqref{eq.infmethod_pseudoorthnormeigvecs} holds, the energy gaps between the ground state and the low-energy excited states of the original Hamiltonian $\hat{H}$ are given by
\begin{equation}
\Delta _{\{ n_a \}} :=\sum _a \Delta _a n_a
\end{equation}
with $n_a \in\{ 0,1,2,\dotsc\}$. Here,
\begin{equation}
\Delta _a :=\frac{N}{S} \varepsilon _a =2A\varepsilon _a \label{eq.infmethod_elementarygaps}
\end{equation}
are the energy gaps created by the quasi-particle excitations $\hat{b}_a'^\dagger$. We can assume that
\begin{equation}
\Delta _1 <\dots <\Delta _A\iff\varepsilon _1 <\dots <\varepsilon _A
\end{equation}
without loss of generality. Then, the energy gap between the ground and first excited states is
\begin{equation}
\Delta =\Delta _1 .
\end{equation}

Notice that our method of calculating the energy gap is not applicable to a first-order transition point because the quasi-particle excitation $\hat{b}_1'^\dagger$ is a fluctuation around the single global minimum of the classical potential $h(\{\bm{m}_a \} )$. Typically, the classical potential $h(\{\bm{m}_a \} )$ is a double-well potential at a first-order transition and the ground and first excited states are superpositions of the states localized at the two minima. To calculate the gap between these states analytically, the discrete WKB or instantonic method would be needed\cite{Garg1998,Garg2003,Bapst2012,Ohkuwa2017}.

\section{Analysis of a System with Sparse Interactions between Subsystems} \label{sec.sinfmethod}

Let us analyze a semi-infinite-range spin system which consists of several subsystems and has sparse interactions between subsystems. We denote the number of spins by $N$, the number of subsystems by $A=\mathcal{O} (N^0)$, and the Pauli operator at site $(a,r)$ ($a=1,\dots ,A$; $r=1,\dots ,N/A$) by $\hat{\bm{\sigma}}_{ar} =(\hat{\sigma}_{ar}^\alpha )_{\alpha =x,y,z}$. Differently from Appendix~\ref{sec.infmethod}, we consider the Hamiltonian in the following form:
\begin{equation}
\hat{H} =Nh_\mathrm{m} (\{\hat{\bm{m}}_a \} )-\frac{1}{2} \sum _{ab} \sum _r \hat{\bm{\sigma}}_{ar} \cdot K_{ab} \hat{\bm{\sigma}}_{br} ,
\end{equation}
where the first term $h_\mathrm{m} (\{\hat{\bm{m}}_a \} )$ is the mean-field part written as a function of $\hat{\bm{m}}_a =\frac{A}{N} \sum _r \hat{\bm{\sigma}}_{ar}$ and the second term is the sum of sparse interactions between the subsystems which are determined by the matrices $K_{ab} =(K_{ab}^{\alpha\beta})_{\alpha ,\beta =x,y,z}$. We assume that $h_\mathrm{m} (\{\hat{\bm{m}}_a \} )$ is a polynomial of degree $P=\mathcal{O} (N^0)$ with coefficients of $\mathcal{O} (N^0)$. We can set $K_{aa}^{\alpha\beta} =0$ and $K_{ab}^{\alpha\beta} =K_{ba}^{\beta\alpha} \in\mathbb{R}$ without loss of generality, because the term with the coefficient matrix $K_{aa}$ can be included in the mean-field part and $[\hat{\sigma}_a^\alpha ,\hat{\sigma}_b^\beta ]=0$ for $a\not= b$.

Now we introduce the path-integral representation of the partition function $Z=\Tr e^{-\beta\hat{H}}$ with inverse temperature $\beta$:
\begin{align}
Z & =\left(\prod _{ar} \int\mathcal{D} \bm{n}_{ar} \right)\exp\left[ -\int _0^\beta d\tau\left(\sum _{ar} \braket{\bm{n}_{ar} (\tau )|\frac{d}{d\tau}|\bm{n}_{ar} (\tau )} \right.\right. \notag \\
& \hphantom{{} = {}} \left.\vphantom{-\int _0^\beta d\tau} \left.\vphantom{\sum _{ar} \braket{\bm{n}_{ar} (\tau )|\frac{d}{d\tau}|\bm{n}_{ar} (\tau )}} +\braket{\{\bm{n}_{ar} (\tau )\} |\hat{H}|\{\bm{n}_{ar} (\tau )\}} \right)\right] . \label{eq.sinfmethod_pathint}
\end{align}
Here, $\ket{\{\bm{n}_{ar} \}} =\bigotimes _{ar} \ket{\bm{n}_{ar}}$ is the product state of the spin coherent states $\ket{\bm{n}_{ar}}$ determined by unit vectors $\bm{n}_{ar}$. The spin coherent state $\ket{\bm{n}_{ar}}$ at each site is the normalized eigenstate of $\hat{\bm{\sigma}}_{ar} \cdot\bm{n}_{ar}$ with the eigenvalue one. In addition, $\mathcal{D} \bm{n}_{ar}$ is the functional measure which is the product of the measures on the 2-sphere over imaginary time $\tau\in [0,\beta ]$.

The energy expectation value in the spin coherent state is given by
\begin{align}
\braket{\{\bm{n}_{ar} \} |\hat{H}|\{\bm{n}_{ar} \}} & =N\braket{\{\bm{n}_{ar} \} |h_\mathrm{m} (\{\hat{\bm{m}}_a \} )|\{\bm{n}_{ar} \}} \notag \\
& \hphantom{{} = {}} -\frac{1}{2} \sum _{ab} \sum _r \bm{n}_{ar} \cdot K_{ab} \bm{n}_{br} . \label{eq.sinfmethod_enexp}
\end{align}
We use an approximation for the first term $\braket{\{\bm{n}_{ar} \} |h_\mathrm{m} (\{\hat{\bm{m}}_a \} )|\{\bm{n}_{ar} \}}$. The mean-field part of the Hamiltonian density $h_\mathrm{m} (\{\hat{\bm{m}}_a \} )$ is a linear combination of $\hat{m}_{a_1}^{\alpha _1} \dotsm\hat{m}_{a_p}^{\alpha _p}$ ($p=1,\dots ,P$), which has the expectation value
\begin{align}
& \hphantom{{} = {}} \braket{\{\bm{n}_{ar} \} |\hat{m}_{a_1}^{\alpha _1} \dotsm\hat{m}_{a_p}^{\alpha _p}|\{\bm{n}_{ar} \}} \notag \\
& =\left(\frac{A}{N} \right) ^p \sum _{r_1 ,\dots ,r_p} \braket{\{\bm{n}_{ar} \} |\hat{\sigma}_{a_1 r_1}^{\alpha _1} \dotsm\hat{\sigma}_{a_p r_p}^{\alpha _p}|\{\bm{n}_{ar} \}} \notag \\
& =\left(\frac{A}{N} \right) ^p \sum _{\substack{r_1 ,\dots ,r_p \\ q\not= q'\Longrightarrow r_q \not= r_{q'}}} \braket{\{\bm{n}_{ar} \} |\hat{\sigma}_{a_1 r_1}^{\alpha _1} \dotsm\hat{\sigma}_{a_p r_p}^{\alpha _p}|\{\bm{n}_{ar} \}} +\mathcal{O} (N^{-1}) \notag \\
& =\left(\frac{A}{N} \right) ^p \sum _{\substack{r_1 ,\dots ,r_p \\ q\not= q'\Longrightarrow r_q \not= r_{q'}}} n_{a_1 r_1}^{\alpha _1} \dotsm n_{a_p r_p}^{\alpha _p} +\mathcal{O} (N^{-1}) \notag \\
& =\left(\frac{A}{N} \right) ^p \sum _{r_1 ,\dots ,r_p} n_{a_1 r_1}^{\alpha _1} \dotsm n_{a_p r_p}^{\alpha _p} +\mathcal{O} (N^{-1}) \notag \\
& =\left(\frac{A}{N} \sum _{r_1} n_{a_1 r_1}^{\alpha _1} \right)\dotsm\left(\frac{A}{N} \sum _{r_p} n_{a_p r_p}^{\alpha _p} \right) +\mathcal{O} (N^{-1}).
\end{align}
In the third and fifth lines of this equation, we used the fact that the number of $(r_1 ,\dots ,r_p)$ including equal indices is of $\mathcal{O} (N^{p-1})$, while the number of $(r_1 ,\dots ,r_p)$ whose elements are different from each other is $\frac{N}{A} \left(\frac{N}{A} -1\right)\dotsm\left(\frac{N}{A} -p+1\right) =\mathcal{O} (N^p)$. Hence, we obtain the simplified expression
\begin{equation}
\braket{\{\bm{n}_{ar} \} |h_\mathrm{m} (\{\hat{\bm{m}}_a \} )|\{\bm{n}_{ar} \}} =h_\mathrm{m} \left(\left\{\frac{A}{N} \sum _r \bm{n}_a \right\}\right) +\mathcal{O} (N^{-1}). \label{eq.sinfmethod_enexpmf}
\end{equation}
We ignore the term of $\mathcal{O} (N^{-1})$, which yields a non-extensive correction to $\braket{\{\bm{n}_{ar} \} |\hat{H}|\{\bm{n}_{ar} \}}$.

Combining Eqs.~\eqref{eq.sinfmethod_pathint}, \eqref{eq.sinfmethod_enexp}, and \eqref{eq.sinfmethod_enexpmf} yields
\begin{align}
Z & =\left(\prod _{ar} \int\mathcal{D} \bm{n}_{ar} \right)\exp\left[ -\int _0^\beta d\tau\left(\sum _{ar} \braket{\bm{n}_{ar} (\tau )|\frac{d}{d\tau}|\bm{n}_{ar} (\tau )} \right.\right. \notag \\
& \hphantom{{} = {}} \left.\left. {} +Nh_\mathrm{m} \left(\left\{\frac{A}{N} \sum _r \bm{n}_a (\tau )\right\}\right)\right.\right. \notag \\
& \hphantom{{} = {}} \left.\left. {} -\frac{1}{2} \sum _{ab} \sum _r \bm{n}_{ar} (\tau )\cdot K_{ab} \bm{n}_{br} (\tau )\right)\right] . \label{eq.sinfmethod_partfuncwom}
\end{align}
We insert the path integral of the delta functional
\begin{align}
1 & =\int\mathcal{D} \bm{m}_a \,\delta\left[\frac{N}{A} \bm{m}_a (\tau )-\sum _r \bm{n}_{ar} (\tau )\right] \notag \\
& =\int\mathcal{D} \bm{m}_a \int\mathcal{D} \tilde{\bm{m}}_a \notag \\
& \hphantom{{} = {}} \exp\left[ -\int _0^\beta d\tau\,\tilde{\bm{m}}_a (\tau )\cdot\left(\frac{N}{A} \bm{m}_a (\tau )-\sum _r \bm{n}_{ar} (\tau )\right)\right]
\end{align}
for $a=1,\dots ,A$ into Eq.~\eqref{eq.sinfmethod_partfuncwom}. Then, we obtain
\begin{align}
Z & =\left(\prod _a \int\mathcal{D} \bm{m}_a \int\mathcal{D} \tilde{\bm{m}}_a \right) \notag \\
& \hphantom{{} = {}} \exp\left[ -N\int _0^\beta d\tau\left(\frac{1}{A} \sum _a \tilde{\bm{m}}_a (\tau )\cdot\bm{m}_a (\tau )+h_\mathrm{m} (\{\bm{m}_a (\tau )\} )\right)\right] \notag \\
& \hphantom{{} = {}} \times\prod _r \left\{\left(\prod _a \int\mathcal{D} \bm{n}_{ar} \right)\right. \notag \\
& \hphantom{{} = {} \times} \exp\left[ -\int _0^\beta d\tau\left(\sum _a \braket{\bm{n}_{ar} (\tau )|\frac{d}{d\tau}|\bm{n}_{ar} (\tau )} \right.\right. \notag \\
& \hphantom{{} = {} \times} \left.\left.\left. {} -\sum _a \tilde{\bm{m}}_a (\tau )\cdot\bm{n}_{ar} (\tau )-\frac{1}{2} \sum _{ab} \bm{n}_{ar} (\tau )\cdot K_{ab} \bm{n}_{br} (\tau )\right)\right]\right\} \notag \\
& =\left(\prod _a \int\mathcal{D} \bm{m}_a \int\mathcal{D} \tilde{\bm{m}}_a \right) \notag \\
& \hphantom{{} = {}} \exp\left[ -N\int _0^\beta d\tau\left(\frac{1}{A} \sum _a \tilde{\bm{m}}_a (\tau )\cdot\bm{m}_a (\tau )+h_\mathrm{m} (\{\bm{m}_a (\tau )\} )\right)\right] \notag \\
& \hphantom{{} = {}} \times\left\{\left(\prod _a \int\mathcal{D} \bm{n}_a \right)\exp\left[ -\int _0^\beta d\tau\left(\sum _a \braket{\bm{n}_a (\tau )|\frac{d}{d\tau}|\bm{n}_a (\tau )} \right.\right.\right. \notag \\
& \hphantom{{} = {} \times} \left.\left.\left. {} -\sum _a \tilde{\bm{m}}_a (\tau )\cdot\bm{n}_a (\tau )-\frac{1}{2} \sum _{ab} \bm{n}_a (\tau )\cdot K_{ab} \bm{n}_b (\tau )\right)\right]\right\} ^{N/A} ,
\end{align}
where we simplified the product over $r=1,\dots ,N/A$ as the ($N/A$)th power, because the factor in the expression of $Z$ depends on $r$ only through $\bm{n}_{ar}$.

Let us evaluate the asymptotic form of the partition function in the thermodynamic limit $N\to\infty$ by the stationary-phase approximation. We assume that the stationary path (i.e., the set of the functions $\bm{m}_a (\tau )$ and $\tilde{\bm{m}}_a (\tau )$ for which the functional derivatives of the integrand of $Z$ vanish) satisfies the static ansatz
\begin{equation}
\bm{m}_a (\tau )=\bm{m}_a ,\quad\tilde{\bm{m}}_a (\tau )=\tilde{\bm{m}}_a .
\end{equation}
Then, we can write the partition function as an ordinary integral with respect to $\bm{m}_a$ and $\tilde{\bm{m}}_a$:
\begin{align}
Z & =\left(\prod _a \int d^3 \bm{m}_a \int d^3 \tilde{\bm{m}}_a \right) \notag \\
& \hphantom{{} = {}} \exp\left[ -N\beta\left(\frac{1}{A} \sum _a \tilde{\bm{m}}_a \cdot\bm{m}_a +h_\mathrm{m} (\{\bm{m}_a \} )\right)\right] \notag \\
& \hphantom{{} = {}} \times\left\{\left(\prod _a \int\mathcal{D} \bm{n}_a \right)\exp\left[ -\int _0^\beta d\tau\left(\sum _a \braket{\bm{n}_a (\tau )|\frac{d}{d\tau}|\bm{n}_a (\tau )} \right.\right.\right. \notag \\
& \hphantom{{} = {} \times} \left.\left.\left. {} -\sum _a \tilde{\bm{m}}_a \cdot\bm{n}_a (\tau )-\frac{1}{2} \sum _{ab} \bm{n}_a (\tau )\cdot K_{ab} \bm{n}_b (\tau )\right)\right]\right\} ^{N/A} .
\end{align}
Replacing the path integral over $\bm{n}_a (\tau )$ with the trace of an exponentiated operator results in
\begin{align}
Z & =\left(\prod _a \int d^3 \bm{m}_a \int d^3 \tilde{\bm{m}}_a \right) \notag \\
& \hphantom{{} = {}} \exp\left[ -N\beta\left(\frac{1}{A} \sum _a \tilde{\bm{m}}_a \cdot\bm{m}_a +h_\mathrm{m} (\{\bm{m}_a \} )\right)\right. \notag \\
& \hphantom{{} = {}} \left.\vphantom{\left(\frac{1}{A} \right)} +\frac{N}{A} \ln\Tr e^{-\beta\hat{h}_\mathrm{eff} (\{\tilde{\bm{m}}_a \} )} \right] . \label{eq.sinfmethod_partfuncst}
\end{align}
We defined the effective Hamiltonian of an $A$-spin system as
\begin{equation}
\hat{h}_\mathrm{eff} (\{\tilde{\bm{m}}_a \} )=-\sum _a \tilde{\bm{m}}_a \cdot\hat{\bm{\sigma}}_a -\frac{1}{2} \sum _{ab} \hat{\bm{\sigma}}_a \cdot K_{ab} \hat{\bm{\sigma}}_b ,
\end{equation}
where $\hat{\bm{\sigma}}_a$ ($a=1,\dots ,A$) are the Pauli operators.

Applying the saddle-point method (the stationary-phase approximation) to Eq.~\eqref{eq.sinfmethod_partfuncst} in the thermodynamic limit $N\to\infty$, we derive the expression of the free-energy density $f=-\frac{1}{N\beta} \ln Z$ as follows:
\begin{equation}
f=\frac{1}{A} \sum _a \tilde{\bm{m}}_a \cdot\bm{m}_a +h_\mathrm{m} (\{\bm{m}_a \} )-\frac{1}{A\beta} \ln\Tr e^{-\beta\hat{h}_\mathrm{eff} (\{\tilde{\bm{m}}_a \} )} . \label{eq.sinfmethod_freeenergy}
\end{equation}
We can calculate the order parameters $\bm{m}_a$, which are the magnetizations for the $A$ subsystems, and their conjugate parameters $\tilde{\bm{m}}_a$, by solving the saddle-point equations $\partial f/\partial\bm{m}_a =\partial f/\partial\tilde{\bm{m}}_a =\bm{0}$, i.e.,
\begin{align}
\bm{m}_a & =\braket{\hat{\bm{\sigma}}_a}_{\{\tilde{\bm{m}}_a \}} (\beta ), \label{eq.sinfmethod_sdlptordprm} \\
\tilde{\bm{m}}_a & =-A\frac{\partial h_\mathrm{m}}{\partial\bm{m}_a} . \label{eq.sinfmethod_sdlptconjprm}
\end{align}
Here,
\begin{equation}
\braket{\cdots}_{\{\tilde{\bm{m}}_a \}} (\beta ):=\frac{\Tr e^{-\beta\hat{h}_\mathrm{eff} (\{\tilde{\bm{m}}_a \} )} \cdots}{\Tr e^{-\beta\hat{h}_\mathrm{eff} (\{\tilde{\bm{m}}_a \} )}}
\end{equation}
is the thermal expectation value for the effective $A$-spin system.

Let us denote the eigenvalues of $\hat{h}_\mathrm{eff} (\{\tilde{\bm{m}}_a \} )$ by $\lambda _n (\{\tilde{\bm{m}}_a \} )$ ($n=0,1,\dots ,2^A -1$) and the corresponding eigenvectors by $\ket{\lambda _n}_{\{\tilde{\bm{m}}_a \}}$. Suppose that the eigenvalues $\lambda _n (\{\tilde{\bm{m}}_a \} )$ are sorted in ascending order and the minimum eigenvalue is $g(\{\tilde{\bm{m}}_a \} )$-fold degenerate:
\begin{equation}
\lambda _0 (\{\tilde{\bm{m}}_a \} )=\dots =\lambda _{g(\{\tilde{\bm{m}}_a \} )-1} <\lambda _{g(\{\tilde{\bm{m}}_a \} )} \leq\dots\leq\lambda _{2^A -1} .
\end{equation}
Defining the gaps from the minimum eigenvalue as $\delta _n (\{\tilde{\bm{m}}_a \} )=\lambda _n (\{\tilde{\bm{m}}_a \} )-\lambda _0 (\{\tilde{\bm{m}}_a \} )$, we find that Eqs.~\eqref{eq.sinfmethod_freeenergy} and \eqref{eq.sinfmethod_sdlptordprm} are rewritten as
\begin{equation}
f=\frac{1}{A} \sum _a \tilde{\bm{m}}_a \cdot\bm{m}_a +h_\mathrm{m} (\{\bm{m}_a \} )+\frac{1}{A} \lambda _0 (\{\tilde{\bm{m}}_a \} )-\frac{1}{A\beta} \ln\sum _n e^{-\beta\delta _n (\{\tilde{\bm{m}}_a \} )} \label{eq.sinfmethod_freeenergyeigs}
\end{equation}
and
\begin{equation}
\bm{m}_a =\frac{\sum _n e^{-\beta\delta _n (\{\tilde{\bm{m}}_a \} )} \braket{\lambda _n|\hat{\bm{\sigma}}_a|\lambda _n}_{\{\tilde{\bm{m}}_a \}}}{\sum _n e^{-\beta\delta _n (\{\tilde{\bm{m}}_a \} )}} , \label{eq.sinfmethod_sdlptordprmeigs}
\end{equation}
where $\braket{\lambda _n|\hat{\bm{\sigma}}_a|\lambda _n}_{\{\tilde{\bm{m}}_a \}}$ is the expectation value of $\hat{\bm{\sigma}}_a$ in the state $\ket{\lambda _n}_{\{\tilde{\bm{m}}_a \}}$.

Now we take the zero-temperature limit $\beta\to\infty$. Since the excited states of the effective Hamiltonian $\hat{h}_\mathrm{eff} (\{\tilde{\bm{m}}_a \} )$ do not contribute to the free-energy density~\eqref{eq.sinfmethod_freeenergyeigs} and the saddle-point equation~\eqref{eq.sinfmethod_sdlptordprmeigs} in the limit $\beta\to\infty$, the free-energy density $f$ approaches the ground-state energy density
\begin{equation}
u=\frac{1}{A} \sum _a \tilde{\bm{m}}_a \cdot\bm{m}_a +h_\mathrm{m} (\{\bm{m}_a \} )+\frac{1}{A} \lambda _0 (\{\tilde{\bm{m}}_a \} )
\end{equation}
and the parameters $\bm{m}_a$ and $\tilde{\bm{m}}_a$ satisfy
\begin{equation}
\bm{m}_a =\frac{1}{g(\{\tilde{\bm{m}}_a \} )} \sum _{n=0}^{g(\{\tilde{\bm{m}}_a \} )-1} \braket{\lambda _n|\hat{\bm{\sigma}}_a|\lambda _n}_{\{\tilde{\bm{m}}_a \}} .
\end{equation}
Notice that Eq.~\eqref{eq.sinfmethod_sdlptconjprm} still holds in the zero-temperature limit $\beta\to\infty$.

\section{Results for the Total $XX$ Catalyst} \label{sec.totalcatalyst}

We consider the weak-strong cluster problem with the total $XX$ catalyst, which has both of intercluster and intracluster $XX$ interactions. The Hamiltonian is given by Eq.~\eqref{eq.wscl_mf_hamiltonian} for dense intercluster interactions and Eq.~\eqref{eq.wscl_sp_hamiltonian} for sparse ones. We set $\gamma _1 (s)=\gamma _2 (s)=s$ and $(\xi _{11} ,\xi _{22} ,\xi _{12})=(\xi /2,\xi /2,\xi )$ in both cases. Notice that in the case of the dense intercluster interactions, the $XX$ catalyst is proportional to the total $x$-magnetization operator squared $((\hat{m}_1^x +\hat{m}_2^x)/2)^2$.

\begin{figure*}
\centering
\includegraphics{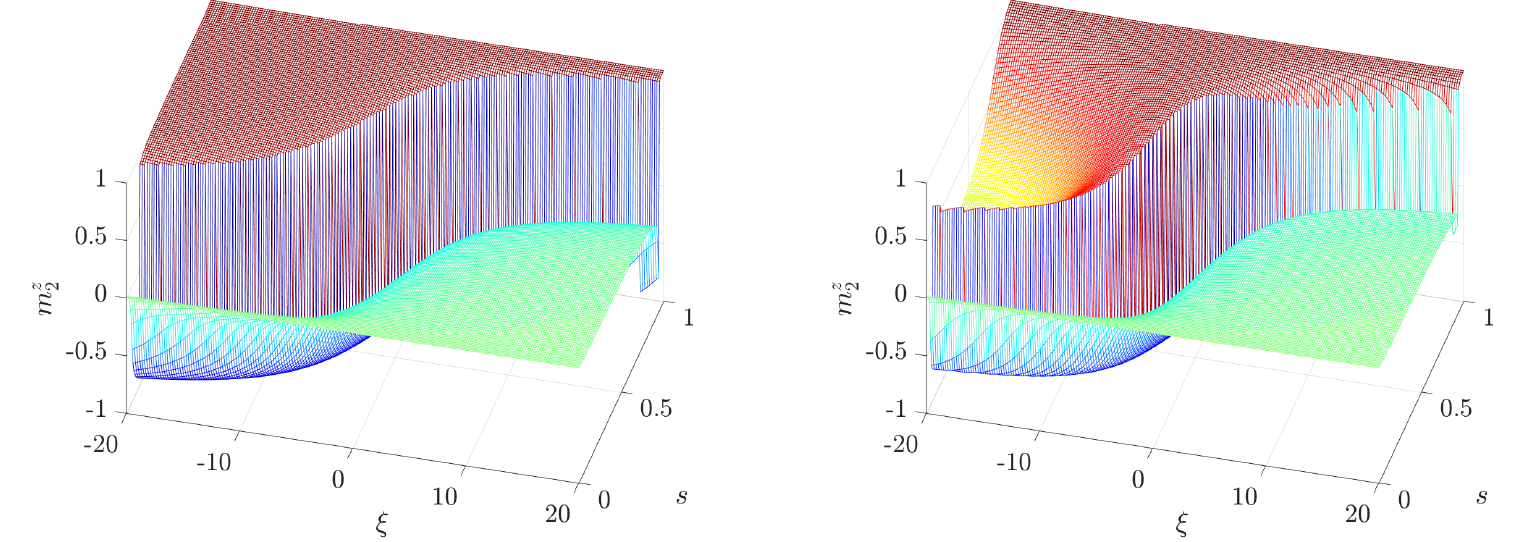}
\caption{(Color online) Magnetization in the weak cluster $m_2^z$ of the weak-strong cluster problem with dense (left) or sparse (right) intercluster interactions for $\gamma _1 (s)=\gamma _2 (s)=s$ and $(\xi _{11} ,\xi _{22} ,\xi _{12})=(\xi /2,\xi /2,\xi )$.} \label{fig.totalcatalyst_mzXXinbt}
\end{figure*}

Figure~\ref{fig.totalcatalyst_mzXXinbt} shows the magnetization in the weak cluster $m_2^z$ for the dense interactions between the clusters and for sparse ones. We find that the total $XX$ catalyst cannot eliminate the first-order transition, whether the intercluster interactions are dense or sparse and whether the catalyst is stoquastic or non-stoquastic. The result in the case of the dense intercluster interactions with the non-stoquastic catalyst $\xi <0$ is consistent with the numerical consequence\cite{Albash2019}.

\bibliography{69558.bib}

\end{document}